\documentclass[a4paper,10pt
  ]
  {article}

\usepackage{hyperref}
\usepackage[dvips]{graphicx}

\def\v#1{{\bf#1}}
\def\be{\begin{equation}}
\def\ee{\end{equation}}
\def\bea{\begin{eqnarray}}
\def\eea{\end{eqnarray}}
\def\ahalf{{\textstyle{1\over2}}}

\newcommand{\bfalpha}{\mbox{\boldmath$\alpha$\unboldmath}}

\newcommand{\bfsigma}{\mbox{\boldmath$\sigma$\unboldmath}}

\newcommand{\bfkappa}{\mbox{\boldmath$\kappa$\unboldmath}}

\def\ie{{\it i.e.\,}}
\def\etal{{\it et al.   }}
\def\ocal{\mbox{$\cal O\,$}}

\def\ecal{\mbox{$\cal E\,$}}
\def\fcal{\mbox{$\cal F\,$}}

\def\<{\langle}
\def\>{\rangle}

\title{ \textbf{The Dirac-Moshinsky Oscillator: Theory and Applications} }

\author{Emerson Sadurn\'i}

\date{Institut f\"ur Quantenphysik, Ulm Universit\"at, Albert-Einstein Allee 11 89081 Ulm - Germany.}

\begin{document}

\maketitle

\begin{abstract}

This work summarizes the most important developments in the construction and application of the Dirac-Moshinsky oscillator (DMO) with which the author has come in contact. The literature on the subject is voluminous, mostly because of the avenues that exact solvability opens towards our understanding of relativistic quantum mechanics. Here we make an effort to present the subject in chronological order and also in increasing degree of complexity of its parts. We start our discussion with the seminal paper by Moshinsky and Szczepaniak and the immediate implications stemming from it. Then we analyze the extensions of this model to many particles. The one-particle DMO is revisited in the light of the Jaynes-Cummings model in quantum optics and exactly solvable extensions are presented. Applications and implementations in hexagonal lattices are given, with a particular emphasis in the emulation of graphene in electromagnetic billiards. \vspace{12pt}
\newline
{\bf PACS\ }: 03.65.Pm, 12.40.Yx, 73.22.Pr, 37.30.+i \newline
{\bf Keywords\ }: Dirac Oscillator, Hadrons, Jaynes-Cummings model, Graphene.

\end{abstract}




\tableofcontents

\section{Introduction}

The harmonic oscillator is the paradigm of integrability and solvability with applications to many branches of physics. As it was written by Moshinsky \cite{moshbook} "...A complete analysis of the subject would require an encyclopedia, within which one 
of the volumes could be (our) book". Is it possible to promote all these features to a relativistic quantum-mechanical model? This is the question that Moshinsky and Szczepaniak answered in a seminal paper more than twenty years ago. Today we can see how this idea has been exploited in several ways by using the Dirac-Moshinsky oscillator (DMO) as a way to understand better the mathematical structure of solvable Dirac equations. But beyond the mathematical developments surrounding this system, in these lecture notes we would like to emphasize that some applications can be found in areas of physics of current interest, such as the study of electrons in two dimensional materials (for example, graphene) and the interaction of atoms with electromagnetic fields in cavities (the Jaynes-Cummings model). The notes are divided in four sections. In the first section we give a detailed introduction to the subject, covering symmetries, Lorentz covariance and algebraic solvability. In section two, we review the many body theory for the Dirac equation (in first quantization)  and the key points of the spectral structure of these systems. We continue with the formulation of the Dirac oscillator as an interaction between Dirac particles and a brief mention to hadronic spectroscopy is made. In section three we present solvable extensions to the single particle DMO in the context of isospin fields and continue with the formulation of an exact mapping of such extensions to quantum optical cavities. Finally, in section four we deal with tight binding lattices, two dimensional systems and the effective Dirac equations appearing in materials such as graphene and Boron Nitride. In the same section we develop the same idea in the context of electromagnetic billiards and a deformation method is proposed, leading to a realization of a DMO in one and two dimensions.

\section{The Dirac-Moshinsky oscillator for one particle}

\subsection{The Dirac Oscillator as proposed by Moshinsky and Szczepaniak}


Our purpose is to review the construction of an interaction for relativistic systems (particles) producing bound states for arbitrarily high energies with analytically solvable spectrum. Lorentz invariance is crucial. This was achieved by Moshinsky and Szczepaniak (1989) \cite{dmo} with further generalizations to describe interacting particles \cite{list} through Poincare invariant equations.

A naive approach to the problem is to propose a one-particle relativistic equation in the form 

\bea
(c^2 \hbar^2 \triangle + m^2 c^4 + \frac{1}{2} m \omega^2 r^2) \phi = 0 
\eea
with the trivial result that the energies become

\bea
E^2 - m^2 c^4 = 2\omega \hbar (n + \frac{3}{2})
\eea
However, the Lorentz invariance of the problem is not clear in this simple picture. It is also necessary to find a first order equation in time (as Dirac originally proposed through his equation \cite{diracbook}) for a good application to initial condition problems, as it is the case for hamiltonian systems in quantum mechanics.

Interactions which are linear in the coordinate were introduced in \cite{ito}, but Moshinsky and Szczepaniak introduced and solved a Dirac equation with a hamiltonian of the form
\bea
H= c \bfalpha \cdot \left(\v p \pm i \omega m \beta \v r \right) + m c^2 \beta
\label{0.1}
\eea
where $\v p = -i \hbar \nabla$ and the Dirac matrices are given by  $\beta=\gamma^{0}$, $\alpha^{i}=\beta \gamma^{i}$, $i=1,2,3$. The $\gamma$'s, in turn, are given in the usual representation

\bea
\gamma_j = \left( \begin{array}{cc} 0 & i \sigma_j \\  i \sigma_j & 0 \end{array} \right), \quad \gamma_0 = \left( \begin{array}{cc} \v 1_2 & 0 \\ 0 & -\v 1_2 \end{array} \right).
\label{uno.1}
\eea
The double sign in the frequency $\omega$ written in (\ref{0.1}) indicates that similar results can be obtained independently of this choice. In this framework, both coordinate and momentum operators must appear in linear
form in order to preserve integrability: a clear indication of phase space symmetry. The symmetry Lie algebra of this system was investigated in \cite{quesne}, and the corresponding generators are now represented in the algebra of Dirac matrices. The corresponding group decomposes naturally into $O(4)$ (compact component representing a non-relativistic oscillator) and
$O(3,1)$ (non-compact component representing states with infinite degeneracy).
To see how these two types of degeneracies appear, let us analyze the stationary solutions of the Dirac equation with the hamiltonian (\ref{0.1}).

\subsection{Stationary solutions}

\begin{figure}[h]
\begin{center}
\includegraphics[scale=0.45]{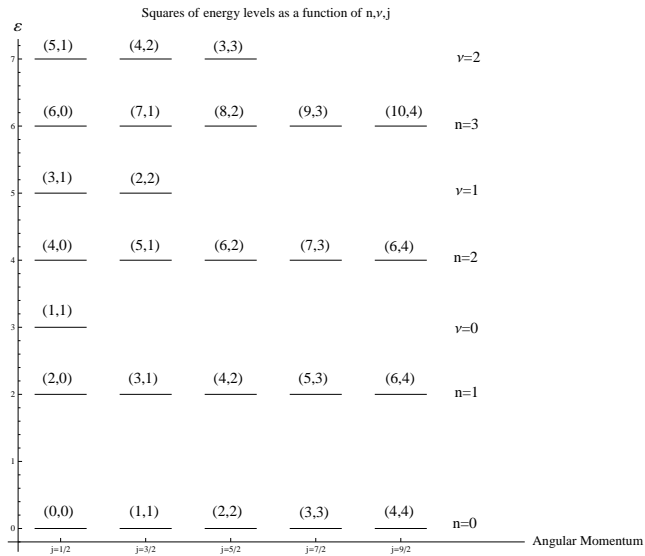}
\end{center}
\caption{Structure of the spectrum for the 3D Dirac-Moshinsky oscillator as proposed by Quesne and Moshinsky \cite{quesne}. The eigenvalue $\epsilon$ (see (\ref{a10.2})) is shown as a function of the total angular momentum and as a function of parity in alternating rows. The quantum number $n \equiv \ahalf(N-j+\ahalf) $ parameterizes the infinitely degenerate states. The number $\nu \equiv \ahalf(N+j-3/2)$ gives the states of finite degeneracy. The nomenclature $(a,b)$ corresponds to $a=2N+l, b=l$ for infinite degeneracy and $a=2\nu+l, b=l$ for finite degeneracy}
\label{figspectrum}
\end{figure}

The stationary form of our equation $H\Psi=E\Psi$ has bispinor solutions of the form

\be
\Psi= \left( \begin{array}{c c} \psi_1 \\ \psi_2 \end{array} \right)
\label{a4}
\ee
satisfying

\be
mc^2 \left(\frac{p^2}{m}+m \omega^2 r^2 + mc^2 -3\omega \hbar -4 \frac{\omega}{\hbar} \v L \cdot \v S \right)\psi_1 = E^2 \psi_1
\label{a5}
\ee

\be
mc^2 \left(\frac{p^2}{m}+m \omega^2 r^2 + mc^2 +3\omega+4\frac{\omega}{\hbar} \v L \cdot \v S \right)\psi_2 = E^2 \psi_2
\label{a6}
\ee
where we have used the customary definition $\v S= \ahalf \hbar \bfsigma$. The wavefunctions are given in terms of the isotropic harmonic oscillator states with total number of quanta $N=2n+l$ and orbital angular momentum $l$. Such states are coupled to the spin $\ahalf$ as we now indicate:

\be
\psi_1= A_{Njl} | N(l,\ahalf)jm \rangle
\label{a8}
\ee

\be
\psi_2= \frac{2c}{\hbar} (E+mc^2)^{-1} \v S \cdot (\v p - i m \omega \v r) \psi_1
\label{a8.1}
\ee
and $A_{Njl}$ is a normalization constant. The energies result in

\bea
E^{2}_{Njl}=m^2c^4+m c^2 \hbar\omega \cases{ 2(N-j)+1  &  $l=j-\ahalf$ \cr  2(N+j)+3 &  $l=j+\ahalf$ }
\label{a9}
\eea
and we write the wavefunctions associated to the positive and negative energies in the form
\be
\Psi_{\pm} = \left( \begin{array}{c c} \psi^{\pm}_1 \\ \psi^{\pm}_2 \end{array} \right), \quad {\rm if} \quad \pm E>0
\label{a10}
\ee
The completeness of these eigenfunctions (\ref{a10}) has been proved in \cite{smyt} as a straightforward exercise. See the figure \ref{figspectrum} for an explanation of the two possibilities of the spectrum according to the parity of the orbital angular momentum $l$. Here it is worth to mention that these solutions constitute a way to write a propagator in spectral form, and that the wavefunctions themselves can be computed through the exact expression of the Dirac oscillator Green's function, obtained in \cite{alhaidari, green}.

\subsubsection{Non-relativistic limit}


Using our previous relations, it is easy to see that

\bea
(E^2-m^2c^4) \psi_1 = \nonumber \\
\left(c^2(p^2+\omega^2 m^2 r^2) -3 \hbar \omega mc^2 -4\frac{\omega}{\hbar} mc^2 \v L \cdot \v S \right)\psi_1 
\label{a10.1}
\eea
for which the relativistic energy given by $\epsilon = E-mc^2 \ll mc^2$ leads to

\bea
\epsilon \psi_1 = \left(H_{HO} -\frac{3}{2} \hbar \omega -2\frac{\omega}{\hbar} \v L \cdot \v S \right)\psi_1 
\label{a10.2}
\eea
where $H_{HO}$ is the usual harmonic oscillator hamiltonian. This shows that the non-relativistic limit reduces to the oscillator without its rest energy (the constant term $-\frac{3}{2} \hbar \omega$ in (\ref{a10.2})) and with a strong spin-orbit coupling. The infinite degeneracy does not disappear, but the negative energy solutions decouple from small components of the spinors as expected. This leaves us with $\psi_1$ as our non-relativistic states.

\subsection{The Dirac-Moshinsky oscillator in Lorentz covariant form}

In relativistic problems, it is always important to formulate everything in Lorentz covariant form. This ensures that the solutions obtained in a particular frame of reference (such as the expressions obtained before) are valid in other inertial frames under the appropriate Lorentz transformations. For our purposes and in what follows, it is convenient to rescale all quantities such that our units give $\hbar=c=1$. Furthermore, let us work with $\omega=1$ and leave the mass $m$ as the only free scale of our system. The Lorentz covariant wave equation for the DMO can be given as

\be
\left(\gamma^{\mu}\left[ p_{\mu}-i r_{\perp \mu} u_{\nu}\gamma^{\nu} \right]+m \right)\Psi=0
\label{a1}
\ee
where $\gamma^{\mu}$ are the Dirac matrices as defined in (\ref{uno.1}) and the perpendicular projection of vectors is given by

\be
r_{\perp \mu}= r_{\mu}-(r^{\nu}u_{\nu})u_{\mu},
\label{a2}
\ee
the vector $u_{\nu}$ being a time-like four vector such that $(u_{\nu})=(1,0,0,0)$ for some inertial frame. In such a frame of reference, (\ref{a1}) can be written as

\be
H\Psi=i\frac{\partial \Psi}{\partial t}
\label{a3}
\ee
with $H$ given by (\ref{0.1}). It is tempting to regard the 4-vector potential
$-i r_{\perp \mu} u_{\nu}\gamma^{\nu}$ as a minimal coupling with a gauge field; however, we must warn the reader that the matrix $\beta$ not only precludes this possibility, but also ensures that such a minimal substitution is not a "pure gauge" interaction, therefore giving non-trivial results. We can give a physical meaning to the vector $u_{\nu}$, as we show in the next section in connection with anomalous coupling.

\subsubsection{Pauli coupling}

The anomalous (Pauli) coupling is a way to introduce interactions with an external field (say, a magnetic field $\v B$) producing terms of the form $\v S \cdot \v B$ in the hamiltonian. It is not our purpose to delve into the nature of such a coupling, but we may emphasize that it provides a possibility of preserving gauge invariance other than the usual minimal coupling.
Using the definition of the spin tensor $S_{\mu \nu} = (1/4)\{ \gamma_{\mu}, \gamma_{\nu} \}$, the Dirac equation for our DMO can be written as

\bea
[\gamma_{\mu}p^{\mu} + m + S_{\mu \nu}F^{\mu \nu}] \psi = 0
\label{uno.2}
\eea
with the choice $F^{\mu \nu}= u^{\mu}r^{\nu}-u^{\nu}r^{\mu}$. The meaning of the external field $F$ can be found by noting that

\bea
\partial_{\mu}F^{\mu \nu} = -u^{\nu},
\label{tres}
\eea
i.e. the vector $u^{\nu}$ can be interpreted as a current. The Maxwell equations for a field given by the tensor $F$ suggest that a constant current given by the r.h.s. of (\ref{tres}) would produce a Dirac oscillator by means of anomalous couplings. Finally, in the frame of reference $(1,0,0,0)$ our current gives
a uniform charge density filling the space.

\subsubsection{The supersymmetric formulation and its extensions} 

A supersymmetric algebra \cite{castanos} which has the squared hamiltonian as its center can be identified as the responsible for the infinite degeneracy of the DMO. We will revisit this point in further sections using a different notation. For now, let us recall what has been done in the context of supersymmetry. When a non-abelian vector potential is used to produce a Dirac oscillator (i.e. $\v p \mapsto \v p + i\beta \v A(\v r)$), one can prove the relations 

\bea
 \{Q_a,Q_b\} = \delta_{ab} (H^2-1), \quad [Q_a,H^2] = 0
\label{super}
\eea
with
\bea
Q_1 = \left( \begin{array}{cc} 0 & \bfsigma \cdot \v a^{\dagger} \\  \bfsigma \cdot \v a & 0 \end{array} \right), \quad Q_2 = \left( \begin{array}{cc} 0 & -i\bfsigma \cdot \v a^{\dagger} \\  i\bfsigma \cdot \v a & 0 \end{array} \right).
\label{charges}
\eea
This structure reveals that more than one choice for the vector potential $\v A(\v r)$ allows analytical solvability: For a general expression of the form $\v a = \v p + iG(r) \v r$, the radial function $G(r)$ may lead to a harmonic oscillator or a Coulomb problem, both of them with additional centrifugal barriers. This is related to the factorization method devised by Infeld and Hull \cite{infeld}, as it was noted in \cite{castanos} in connection with the radial equation resulting from the substitution of $\v a$ in the Dirac equation. One has the radial equation

\bea
\left(G(r)-\frac{l+1}{r} - \frac{d}{dr} \right) \left( G(r)-\frac{l+1}{r} + \frac{d}{dr} \right) R_{N l} = \epsilon R_{N l}
\label{radialsuper}
\eea
and the choices $G(r)=a/r + b, a'/r + b'r$ are possible, leaving the supersymmetry algebra intact. In order to break the degeneracies one may try several tricks. In particular, the introduction of interactions depending explicitly of the total angular momentum is a way of breaking such degeneracies by hand. It is also evident that this approach is attached to the dimensionality of the problem since the radial equation has been used to propose the corresponding extensions. In the following, we shall use an alternative approach to understand infinite degeneracies in connection with dimensionality (2 or 3 dimensions). It will result that the one-dimensional Dirac oscillator admits a superalgebra similar to the one given above, but its degeneracy (if any) is strictly finite.

\subsection{Hilbert space and algebraic structure}

Here we introduce a notation and some concepts which cast the DMO as a bilinear form in bosonic and fermionic operators. This will prove useful in the discussion of invariants and spectral properties. The Lorentz group is locally isomorphic to $SU(2) \times SU^*(2)$. The Hilbert space of our problem is therefore $L_2(C) \times S_3 \times S_3$, where $S_3$ is the space of normalized complex vectors of two entries (Pauli spinors). Obviously, each $S_3$ is a three-sphere \cite{ryder}. In the following, our hamiltonian will be given by

\bea
H= \bfalpha \cdot \left(\v p + i \beta \v r \right) + m \beta
\label{0.0.1}
\eea
with the following representation of the Dirac matrices

\bea
\bfalpha = \left(\begin{array}{cc} 0 & i\bfsigma \\ -i\bfsigma & 0 \end{array} \right), \qquad \beta = \left(\begin{array}{cc} \v 1_2 & 0 \\ 0 & - \v 1_2 \end{array} \right).
\label{0.2}
\eea
For reasons that will become apparent in further sections, we may refer to this representation as quantum-optical. With this notation we may introduce the concept of $*-$spin through the vector $\Sigma_i$, whose z-projection eigenvalues account for big and small components of spinors. Upon rotations, this projection also gives solutions with positive and negative energies.

\bea
\Sigma_+ = \left(\begin{array}{cc} 0 & \v 1_2 \\ 0 & 0 \end{array} \right) = \sigma_+ \otimes \v 1_2, \qquad \Sigma_- = (\Sigma_+)^{\dagger}, \qquad \Sigma_3=\beta
\label{0.3}
\eea
The Hamiltonian can be written in algebraic form as

\bea
H= \Sigma_+ \v S \cdot \v a + \Sigma_- \v S \cdot \v a^{\dagger} + m \Sigma_3, 
\label{0.5}
\eea
The dependence of $H$ on ladder operators makes evident the fact that the following operators are invariant: $I= \v a ^{\dagger} \cdot \v a + \frac{1}{2} \Sigma_3$, $I'= (\v a \cdot \bfsigma )^{\dagger} (\v a \cdot \bfsigma ) + \frac{1}{2} \Sigma_3$. With the integrals of the motion given by a combination of fermionic and bosonic operators, we can obtain the solutions of the eigenvalue problem as follows.

Two states with angular momentum $j$ and satisfying the eigenvalue equation $I |\quad \>=(2n+j-1)|\quad \>$ are given by

\bea
 |\phi_1 \> = |n, (j-1/2,1/2) j, m_j\> |-\>, \quad |\phi_2 \>=|n-1,(j+1/2,1/2) j,m_j\> |+\>.
\label{0.7.1}
\eea
Another pair of states with the same angular momentum $j$ but with eigenvalue $I |\quad
 \>=(2n+j)|\quad \>$ is

\bea
 |\phi_3 \>=|n, (j+1/2,1/2) j, m_j\> |-\>, \quad |\phi_4 \>=|n-1,(j-1/2,1/2) j,m_j\> |+\>.
\label{0.7.2}
\eea
The $2\times 2$ blocks of $H$ obtained from these states can be obtained easily. Here we give such blocks

\bea
 H(j,2n+j-1)= \left(\begin{array}{cc} -m & \sqrt{2n} \\ \sqrt{2n} & m \end{array} \right), 
\label{0.7.3}
\eea

\bea
H(j,2n+j)= \left(\begin{array}{cc} -m & \sqrt{2(n+j)} \\ \sqrt{2(n+j)} & m \end{array} \right).
\eea
From the eigenvalue equation applied to these subspaces, we obtain the well known energies $E^2=m^2+2(n+j)$ and $E^2=m^2+2n$, which correspond to the expressions we have found before.
Infinite and finite degeneracies come from these two blocks respectively. Let us now go further and write similar expressions for low-dimensional Dirac oscillators.

\subsubsection{ Boson-Fermion algebra for $1+1$ and $2+1$ dimensions}

The discussion on the algebraic structure above can be implemented directly in $1+1$ and $2+1$ space-times. For this we have to find the boson (harmonic oscillator) and fermion ($*$-spin) operators which parallel our previous discussion. For the $1+1$ case we define $a,a^{\dagger}$ in terms of the position $x$ and the momentum $p$  in the standard form. For the $2+1$ case, it is useful to define the following chiral creation and annihilation operators (subindex $r$ for right and $l$ for left)

\bea
a_r = a_1 + i a_2, \qquad a_l = a_1 - ia_2 = (a_r)^{*}
\label{0.8}
\eea
with the properties 

\bea
[a_r,a_l]=[a_r,(a_l)^{*}]=0, \quad [a_r,a^{\dagger}_r]=[a_l,a^{\dagger}_l]=4. 
\eea

The low dimensional hamiltonians are

\bea
H^{(1)}=\alpha_1 \left( p+i\beta x \right) + m\beta,
\label{0.9}
\eea
with $\alpha_1 = -\sigma_1, \beta=\sigma_3$ and
\bea
H^{(2)}= \sum_{i=1,2}\alpha_i(p_i+i\beta r_i) +  m \beta,
\label{0.10}
\eea
with the low dimensional Dirac matrices chosen as $\alpha_1 = -\sigma_2, \alpha_2 = - \sigma_1, \beta=\sigma_3$. These hamiltonians can be written in algebraic form as

\bea
H^{(1)}=\sigma_+ a + \sigma_- a^{\dagger} + m \sigma_3
\label{0.11}
\eea

\bea
H^{(2)}=\sigma_+ a_r + \sigma_- a^{\dagger}_r + m \sigma_3
\label{0.12}
\eea
Both of them have a $2\times2$ structure: The spin is absent in one spatial dimension and $\sigma_{\pm}$ corresponds to $*-$spin, while in two dimensions $\sigma_3$ also generates the $U(1)$ spin leading to the total angular momentum $L_3 + \ahalf \sigma_3$. The solvability can be viewed again as a consequence of the existence of invariants. In this case we have

\bea
I^{(1)}= a^{\dagger} a + \frac{1}{2} \sigma_3
\label{0.12.1}
\eea
\bea
I^{(2)} = a_r a^{\dagger}_r + \frac{1}{2} \sigma_3, \qquad J_3 = a_r a^{\dagger}_r - a_l a^{\dagger}_l +\frac{1}{2} \sigma_3 
\label{0.12.2}
\eea
The two dimensional case exhibits some peculiarities. The conservation of angular momentum $J_3$ comes
from the combination of $\bfsigma$ and $a_r$ in $H^{(2)}$,
together with the absence of $a_l, a^{\dagger}_l$. This absence is also responsible for the infinite degeneracy
of all levels. On the other hand, the three dimensional example is manifestly invariant under rotations
due to its dependence on $ \v S \cdot \v a$ and $\v S \cdot \v a^{\dagger} $ and 
its infinite degeneracy comes from the infinitely degenerate operator $(\bfsigma \cdot \v a)(\bfsigma \cdot \v a)^{\dagger}$.

Let us summarize the material of this section. We have learned that the integrability of the harmonic oscillator can be implemented in the context of the Dirac equation by recognizing that coordinates and momenta should lie on an equal footing - the essence of phase space symmetry. The eigenfunctions and energies were given explicitly. A Lorentz covariant equation with a Dirac oscillator potential could be written and interpreted in terms of anomalous coupling and a constant external current. The infinitely degenerate part of the spectrum could be understood either in terms of a supersymmetric algebra (in the $3+1$ dimensional case) or as a consequence of the non-compact part of the symmetry Lie algebra (unitary representations are infinite dimensional). We went further and gave a description of the Dirac oscillator in terms of fermionic and bosonic ladder operators (the operators $a,a^{\dagger}$ and $\sigma_{\pm}$), showing thus the existence of integrals of the motion in a more transparent way. It was also shown that the degeneracies in the $3+1$, $2+1$  and $1+1$ dimensional examples obey a different pattern; in three and two dimensions the parity plays an important role (absence of $j$ and absence of left chiral operators, respectively), while the one-dimensional DMO cannot have infinite degeneracy in despite of the existence of a supersymmetric algebra (one spatial degree of freedom is insufficient).

\section{The many body Dirac equation}

The success of Moshinsky's work related to the harmonic oscillator
of arbitrary particles and dimensions is due to the fact that the results
provided a good basis to solve variational problems in bound
composite systems \cite{moshbook}. This was implemented in composite models describing atomic nuclei. The idea is to extend this success to relativistic quantum mechanics, with the obvious application to many-particle systems where high energies are involved. Many of these examples can be found in the context of hadron physics, where "relativized" models have been proposed \cite{isgur, bjiker}. However, we
need a model which allows the integrability and solvability we are seeking for, in order to understand the structure of multiparticle relativistic formulations, rather than just fitting the results to experimental data. To this end, we start here with the many body Dirac equation as proposed in \cite{list}, followed by the study given by Moshinsky \cite{nikitin}, \cite{sadurni} regarding the positive part of the spectrum and degeneracies in a general framework (the Foldy-Wouthuysen transformation \cite{fw}, \cite{devries}, \cite{nik1}). Then, we present the two and three particle Dirac oscillators as proposed again in the list of works \cite{list}.

\subsection{Poincar\'e invariance of the many body problem}

We review the generalization of the Dirac equation for a system of many particles (carefully treated in \cite{moshbook}). The main idea is to mimic the treatment of many particles in non-relativistic quantum mechanics as the direct product of operator spaces for particle 1,2,..,n. The main equation is defined such that, in the frame
of reference where the center of mass is at rest, we recover a hamiltonian of the form

\bea
H= \sum_{i}^{N} H_i + V(\v x_1, ... , \v x_N)
\label{p31}
\eea
where $H_i$ is the Dirac hamiltonian of the $i$-th particle. The potential $V$ is assumed to be independent of the
center of mass. Such an equation is

\bea
\left[ \sum_{s=1}^{N} \Gamma_s (\gamma_s^{\mu} p_{\mu s} + m_s + \Gamma_s V(x^{s}_{\perp}) ) \right] \psi = 0.
\label{p32}
\eea
The relative coordinates and the time-like relative coordinates are given respectively by

\bea
x^{st}_{\mu}=x^{s}_{\mu}-x^{t}_{\mu}, \quad x^{st}_{\perp \mu}=x^{st}_{\mu}-x^{st}_{\tau} u^{\tau} u_{\mu},
\label{32.1}
\eea
The meaning of the time-like vector defining our preferred frame of reference is obvious, as the hamiltonian stands for the energy at the center of mass with four vector $ P_{\mu}$. We must use the time-like unit vector in the form

\bea
u_{\mu}=(-P_{\tau}P^{\tau})^{-1/2} P_{\mu}.
\label{p33}
\eea
For convenience we have defined the matrices

\bea
\Gamma = \prod_{r=1}^{N} \gamma_{r}^{\mu} u_{\mu}, \qquad \Gamma_s = (\gamma_{s}^{\mu} u_{\mu})^{-1} \Gamma.
\label{p34}
\eea

Taking $P^{i}=0$ and $H=P^{0}$ in (\ref{p32}), one recovers (\ref{p31}).

\subsection{The Foldy-Wouthuysen transformation}

The problem of positive and negative energies in the Dirac equation for arbitrary potentials appeared from the very beginning \cite{diracbook} and it was treated systematically by Foldy and Wouthuysen \cite{fw}, \cite{bjorken}. However, in most of cases such a treatment can be carried out only approximately. See, for example, \cite{devries} for a detailed review of the subject. Remarkably, the Dirac oscillator is one of the examples (together with the free case) in which the corresponding transformation can be carried out analytically \cite{moreno}. There exists a unitary operator which transforms the Dirac hamiltonian into
a diagonal operator in spinorial components. In our algebraic language, the transformation finds the basis
in which the z component of $*$-spin gives the positive and negative energies
of the system. The idea is to
express the hamiltonian in terms of its even part (diagonal matrices) and odd part (anti-diagonal matrices) and find a hermitian operator $S$ such that

\bea
H_{FW}= e^{iS} H_{D} e^{-iS} = \rm{even}.
\eea
For the free particle one has $iS=\beta (\alpha \cdot \v p) \theta$, $\tan (2 \theta \alpha \cdot \v p) = \alpha \cdot \v p$ and

\bea
H_{FW} = \beta \sqrt{p^2+m^2}
\label{fwfree}
\eea
For the three dimensional DMO we use the definition $\alpha \cdot \pi$ as the kinetic energy of the DMO. One has now the relations $iS=\beta (\alpha \cdot \pi) \theta$, $\tan (2 \theta \alpha \cdot \pi) = \alpha \cdot \pi$ and
\bea
H_{FW}=\beta  \sqrt{p^2 + r^2 +(3+2 \v L \cdot \bfsigma) \beta + m^2}.
\label{fwDMO}
\eea
In the following we give a more detailed treatment dealing with arbitrary potentials, first for one particle and then for many particles. This will be useful in our interpretation of a many particle Dirac equation based on the direct product of particle spaces.

\subsection{The many body Foldy-Wouthuysen transformation}

With the aim of characterizing the spectrum of a multibody system with interactions, we seek for an expansion of $H$
in terms of inverse powers of the rest mass. Such an expansion should allow the identification of positive and negative
energies of the model. For one particle in a potential $V$, we have

\bea
H= \ocal + \ecal + V, \quad \ocal = \bfalpha \cdot \v p, \quad \ecal = m \beta.
\label{p35}
\eea
We apply a unitary operator $U=\exp(iS) \exp(iS') \exp(iS'')$,

\bea
\nonumber S = \frac{-i \beta}{2m}\ocal, \quad S' = \frac{-i \beta}{2m}\ocal', \quad S''= \frac{-i \beta}{2m}\ocal'' \\  \ocal' =\frac{\beta}{2m}[\bfalpha \cdot \v p,V], \quad \ocal''= \frac{- (\bfalpha \cdot \v p) p^2}{3m^2}, \quad H'= UHU^{\dagger}
\label{p36}
\eea
Expanding up to $1/( mass )^3$ in the kinetic energy, $1/( mass )^2$ in the potential, we have

\be\begin{array}{l}
                H'=\hat H + V, \; \hat H = \beta \bigg(m+\frac{p^2}{2m}
                - \frac{p^4}{8m^3}\bigg)\\ + \frac{1}{4m^2} \v s
                \cdot \bigg[ (\v p \times \v E) - (\v E\times \v
                p)\bigg] + \frac{1}{8m^2} \nabla^2 V
                        \end{array}\label{mk26}
        \ee
with $\v E= -\nabla V$, $\v S= \frac{-i}{4} \bfalpha \times \bfalpha $. For two particles we define the corresponding matrices as

\bea
\alpha_1= \alpha \otimes \v 1, \qquad  \alpha_2= \v 1 \otimes \alpha \\
\beta_1= \beta \otimes \v 1, \qquad  \beta_2= \v 1 \otimes \beta.
\label{product}
\eea
The hamiltonian is $H=H_1+H_2+V(\v r_1,\v r_2)$. Applying successively $U_1=\exp(iS_1)$ and $U_2=\exp(iS_2)$ one gets

\bea
U_2 U_1 H ( U_2 U_1 )^{\dagger} = \hat H_1 + \hat H_2 + V + { \rm higher \ } \quad { \rm order \ } 
\label{p37}
\eea
In the general case with $n$ particles, one has

\bea
H= \sum_{i=1}^{N} H_i + V(\v r_1,...,\v r_N)
\label{p38}
\eea

\bea
H'= U_N...U_1 H (U_N...U_1)^{\dagger} =\sum_{i=1}^{N} \hat H_i + V(\v r_1,...,\v r_N)
\label{p39}
\eea
with

\be\begin{array}{l}
        \hat H_t = \beta_t \bigg(m_t+ \frac{p^2_t}{2m_t} -
        \frac{p^4_t}{8m_t ^3}\bigg) + \frac{1}{4m_t^2} \v s_t \cdot (\v p_t \times \v E_t-\v
        E_t \times \v p_t ) + \frac{1}{8m_t ^2}
        \nabla^2_t V,\\ t=1,2,\cdots n\end{array}
        \label{mk37}
        \ee
At the end, we have an expression which shows the first relativistic corrections to the kinetic energy, the spin-orbit couplings and the Darwin term \cite{dar}. But the most important feature of the result is that the positive energies can be extracted immediately by selecting the positive eigenvalues of all the $\beta_t$ multiplying the kinetic energy.

\subsubsection{The cockroach nest: Extraordinary infinite degeneracy}

Before completing our task of generalizing the Dirac oscillator to many particles, it is important to understand first the types of degeneracies involved in a multiparticle Dirac equation. Here we show a very simple example.
For commuting Dirac hamiltonians one expects that the total FW transformation can be decomposed
into individual factors corresponding to each hamiltonian. According to our definition of the multiparticle 
FW transformation, the free case gives 

\bea
H_{FW}= \sum_{i=1}^{N} \beta_i \sqrt{p_i^2+m_i^2}
\eea
where it becomes evident that the energies are now added with 'wrong' signs due to the $\beta$ matrices. This
means that the transformation to even hamiltonians contains both particle and anti-particle solutions without
a correction of the signs in front of their kinetic energies. Specifically, for two particles of equal mass described by an observer at the center of mass, only the relative momentum $p$ appears. One of the corresponding energy eigenvalues has the form $\sqrt{p^2+m^2}-\sqrt{(-p)^2+m^2} \equiv 0 $ for any $p$. Moreover, when the relative momentum vanishes the rest energy of the system becomes $0$ instead of the usual value of $2m$. This result seems to be unphysical. Therefore, one has to project the final result onto the purely
positive component, otherwise one would obtain an extraordinary infinite degeneracy \cite{leykoo}.

\subsubsection{Application to the two body problem}

As a point of comparison and before dealing with integrable problems, let us consider now a system of two particles with an interaction given by a quadratic potential. The system is not integrable. The transformed hamiltonian can be approximated by

\be\begin{array}{l}  H' = (\beta_1+\beta_2) \bigg(m+\frac{p^2}{2m}
                - \frac{p^4}{8m^3}\bigg) +V\\+ \frac{1}{4m^2}\bigg( \v
                s_1+\v s_2\bigg)
                \cdot \bigg[ (\v p \times \v E) - (\v E\times \v
                p)\bigg] + \frac{1}{4m^2} \nabla^2 V\end{array}\label{mk40}\ee
with $p$ the magnitude of the relative momentum. The potential is so far arbitrary. We may propose a quadratic interaction $V=\frac{1}{2}m \omega^2 (\v r_1- \v r_2)^2 $.

In the center of mass frame, the choice of positive energy components reduces the hamiltonian to

\be  H' = \bigg(2m+ 3\frac{\omega^2}{8m}\bigg)+\bigg(\frac{p^2}{m}
                 +\frac{m\omega^2r^2}{4}+ \frac{\omega^2}{4m} \v
                S\cdot \v L\bigg)- \frac{p^4}{4m^3}\label{mk47}\ee
with $\v r, \v p$ the relative coordinate and momentum, respectively. The spectrum of the problem is found by diagonalizing
the matrix with elements given by

\be\begin{array}{c} \< n' l,\bigg(\frac12\frac12\bigg) S; j, m|H'|n
l,\bigg(\frac12\frac12\bigg) S, j,
m \>\\=\bigg(2m+\frac{3\omega^2}{8m}+\omega\bigg(2n+l+\frac32\bigg)+
\frac{\omega^2}{8m}[j(j+1)-l(l+1)-s(s+1)]\bigg)\delta_{n
n'}\\
-\frac{1}{4m^3} \< n' l'|p^4|n l \> \end{array}\label{mk474}\ee
where we use two-particle harmonic oscillator states with spin, \ie

\be|n l,\bigg(\frac12\frac12\bigg) S; j, m>\equiv
        \sum_{\mu,\sigma}<l
\mu, S \sigma|j m \>|n l \mu \>|\bigg(\frac12\frac12\bigg) S
    \sigma \>\label{mk473}\ee
We take $N \leq N_{max}$ to get a finite matrix.

As an application, one can describe the mass spectrum of binary systems such as bottomonium or charmonium. It is very important to comment on the flavor of the wavefunctions: According to the theory of particles composed by quarks (Hadrons), one has to use appropriate wavefunctions containing the information of quark flavor and color, the interaction being flavor-blind \cite{blind}. In other words, our potential is permutationally invariant. Moreover, for approximately degenerate quark masses (for example $u$ and $d$ quarks) the states must lie in isospin doublets and should be properly (anti)symmetrized according to the representations of the permutation group \cite{hammermesh}. However, for quarkonia we only have one pair made of quark-antiquark, and the process is trivial. It is therefore sufficient to consider products of wavefunctions of the form Flavor$\times$Spinor which are symmetric, since the colorless feature of the composite demands an antisymmetric color part. In the following, all symmetric and antisymmetric solutions of the eigenvalue problem related to our hamiltonians are considered.

It is possible to introduce quartic corrections to the potential in order to obtain more realistic spectra
 $V'=-\frac{am\omega^4 r^4}{16}$. The
FW transformation of such a term yields next order corrections, therefore we neglect them. The coupling constants
and the rest mass are taken as adjustable parameters. They are fitted to experimental data \cite{pdg} using least dispersion. See the figures. 

\begin{figure}[h]
\begin{center}
\includegraphics[width=7.5cm]{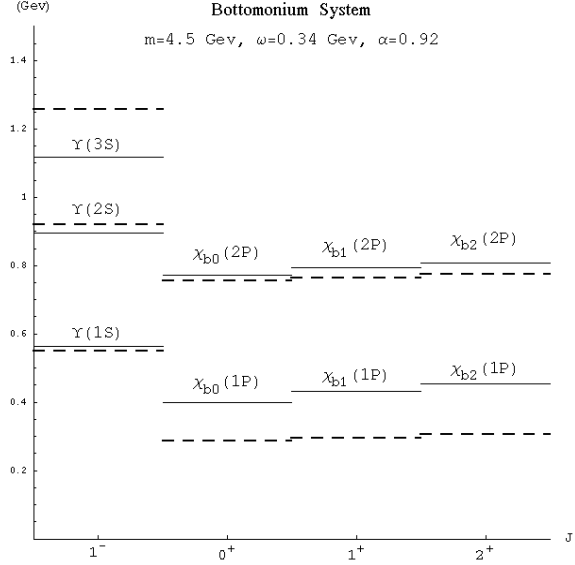} 
\end{center}
\caption{Energy comparison. Solid line: experiment. Dashed line: theory. In $J^P$, $J$ denotes the total angular momentum (also referred to as total spin) and $P$ the parity of the corresponding energy state}
\label{bott}
\end{figure}

\begin{figure}[h]
\begin{center}
\includegraphics[width=7.5cm]{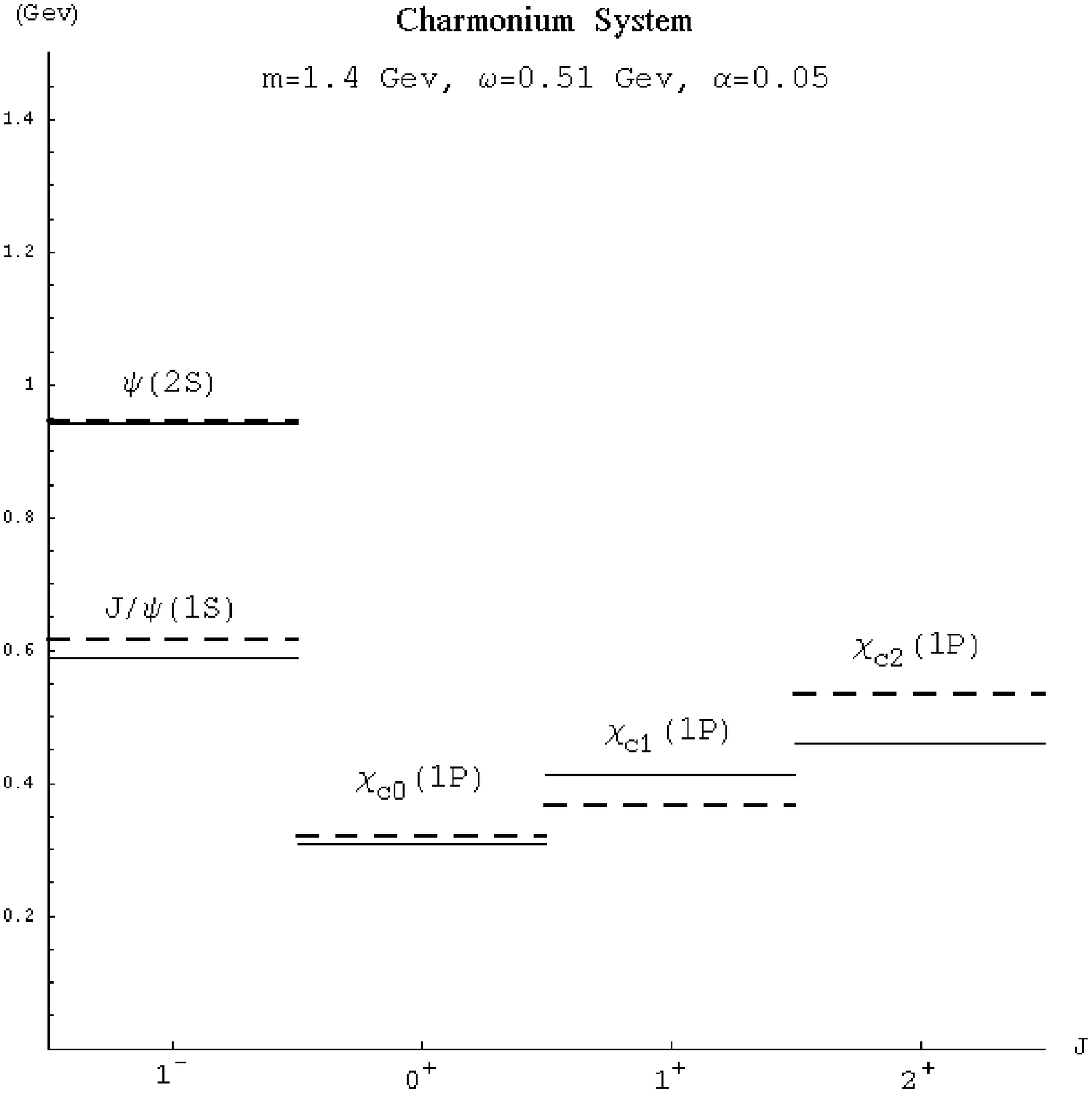}
\end{center}
\caption{Energy comparison. Solid line: experiment. Dashed line: theory. In $J^P$, $J$ denotes the total angular momentum (also referred to as total spin) and $P$ the parity of the corresponding energy state}
\label{charm}
\end{figure}

\subsubsection{Application to the three body problem}

The hamiltonian is now

\be
        H' = \sum^3_{t=1} \beta_t \bigg(m_t+ \frac{p^2_t}{2m_t}  -
        \frac{p^4_t}{8m^3_t}\bigg) + \frac{1}{4m^2_t} \v s_t \cdot (\v p_t \times \v E_t-\v
        E_t \times \v p_t ) + \frac{1}{8m^2_t}
        \nabla^2_t V + V 
       \label{es01}
        \ee

with a flavor-blind potential

\be
V = \frac{M \omega^2}{6} \bigg[ \left(\v r_1 - \v r_2 \right)^2+\left(\v r_2 - \v r_3 \right)^2+\left(\v r_3 - \v r_1 \right)^2 \bigg].
\label{es03}
\ee
Now we use the definition of Jacobi coordinates in order to separate the contribution from the center of mass of the system and the relative coordinates. The harmonic oscillator for $n$ particles with hamiltonian

\be
H= \ahalf \sum_{i=1}^{n} p_i^2 + \frac{\omega^2}{2n}\sum_{i,j=1}^{n}(\v r_i-\v r_j)^2
\label{2.14}
\ee
can be decoupled into $n-1$ oscillators by using the Jacobi coordinates in the form

\bea
\nonumber (\dot p_s)_j &=& [s(s+1)]^{-1/2}\sum_{t=1}^{s}\left(  (p_t)_j -(p_{s+1})_j \right), s=1,\dots,n-1, \\  (\dot p_n)_j &=& n^{-1/2} \sum_{t=1}^{n} (p_t)_j.
\label{2.15}
\eea
Considering that the center of mass is at rest, we obtain

\bea
\nonumber H' &=& Mc^2 + \left( \frac{1}{4}\dot{p}^2_1 + \frac{1}{12}\dot{p}^2_2 \right) \left( \frac{1}{m_1}+\frac{1}{m_2} \right) 
+\frac{1}{3m_3}\dot{p}^2_2+ \frac{1}{\sqrt{12}}\dot{p}_{12}\left( \frac{1}{m_1}-\frac{1}{m_2} \right)
\\ \nonumber &-& \frac{1}{8m_1^3 c^2} \left( \frac{1}{4}\dot{p}^4_1+\frac{1}{36}\dot{p}^4_2+\frac{1}{3}\dot{p}^2_{12}+\frac{1}{6}\dot{p}^2_1 \dot{p}^2_2+\frac{1}{\sqrt{3}}\dot{p}_{12} \dot{p}^2_1+\frac{1}{3\sqrt{3}}\dot{p}_{12} \dot{p}^2_2 \right)
\\ \nonumber &-& \frac{1}{8m_2^3 c^2} \left( \frac{1}{4}\dot{p}^4_1+\frac{1}{36}\dot{p}^4_2+\frac{1}{3}\dot{p}^2_{12}+\frac{1}{6}\dot{p}^2_1 \dot{p}^2_2-\frac{1}{\sqrt{3}}\dot{p}_{12} \dot{p}^2_1-\frac{1}{3\sqrt{3}}\dot{p}_{12} \dot{p}^2_2 \right)
\\ \nonumber &+& \frac{1}{18m_3^3 c^2} \dot{p}^4_2 
\frac{M \omega^2 }{8c^2}\bigg[ \frac{1}{m^2_1}\v S_1\cdot\left( \dot{\v L}_1+\frac{1}{3}\dot{\v L}_2 + \frac{1}{\sqrt{3}}\dot{\v L}_{12} \right)
\bigg] \\ \nonumber  &+& \frac{1}{18m_3^3 c^2} \dot{p}^4_2 
\frac{M \omega^2 }{8c^2}\bigg[ \frac{1}{m^2_2}\v S_2\cdot\left( \dot{\v L}_1+\frac{1}{3}\dot{\v L}_2 - \frac{1}{\sqrt{3}}\dot{\v L}_{12} \right) - \frac{8}{3m^2_3}\v S_3\cdot \dot{\v L}_2 \bigg]
\\ &+&\frac{M \hbar^2 \omega^2}{8c^2}\left( \frac{1}{m^2_1}+\frac{1}{m^2_2}+\frac{1}{m^2_3} \right)+V
\label{es07}
\eea
where $\dot{\v L}_{12} = \dot{\v r_1} \times \dot{\v p_2} + 1 \leftrightarrow 2$ and $\dot{p}_{12} = \dot{\v p}_1 \cdot \dot{\v p}_2$.

The spectrum is obtained by diagonalizing

\bea
\<n'_1,l'_1,n'_2,l'_2,L';\bigg(\frac12\frac12\bigg)T'\frac12S'; j' m'|H'|n_1,l_1,n_2,l_2,L;\bigg(\frac12\frac12\bigg)T\frac12S; j m \>
\label{es09}
\eea
where the states are

\bea \nonumber |n_1,l_1,n_2,l_2,L;\bigg(\frac12\frac12\bigg)T\frac12S; j m \> = \qquad \qquad \qquad \\
       \sum_{\mu,\sigma}<L
\mu, S \sigma|j m \>|n_1,l_1,n_2,l_2,L\mu \>|\bigg(\frac12\frac12\bigg)T \frac12 S
    \sigma>\label{es08}\eea
The matrix elements are computed by means of Racah algebra. We take $N_{max} = 3$. Again, this application does not include other degrees of freedom such as particle flavor. However, this does not preclude the use of these results to obtain a part of the spectrum for a three quark system in which constitutive masses are considered, with the result that the members of our composite system are distinguishable particles (with very different masses and broken degeneracy). As it is evident, the hamiltonian we use in this case is not permutationally invariant, in despite of the fact that the potential enjoys of such a property.

To achieve a better agreement with experimental data, we may introduce a mass and a frequency which depend on the integrals
of the motion.
We include a comparison with the spectra of $\Sigma$ particles (strange baryons). In summary, we have shown that the Poincare invariant equation with arbitrary inter-particle potentials can be treated in the quasi-relativistic approximation by means of the Foldy-Wouthuysen transformation, with the possibility of computing spectra of the transformed hamiltonian. From the computational point of view, the process is not necessarily simple and in despite of the few parameters that can be used to fit energy levels, our understanding of the system is now beyond the symmetry principles underlying the Dirac-Moshinsky oscillator. In the following we describe how to construct the DMO hamiltonian for more than one particle and analyze the corresponding solutions.

\begin{figure}
\begin{center}
\includegraphics[width=9cm]{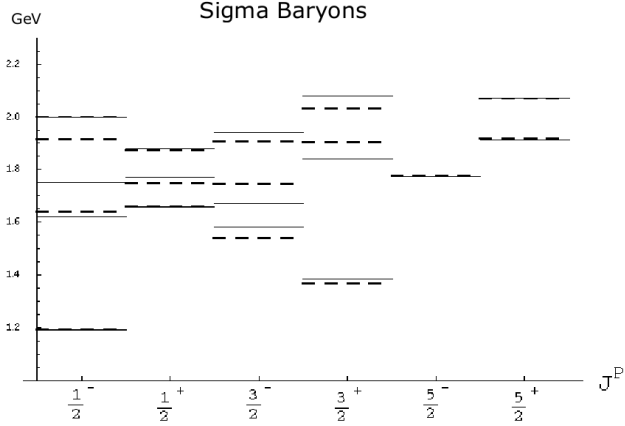} 
\end{center}
\caption{Dashed: Theory. Solid: Experimental. To achieve a better agreement with experiment, we have considered a slight variation of the parameters $\omega, M$ as functions of the integrals of the motion $J^P$ (see figures \ref{bott}, \ref{charm} for the meaning of this nomenclature). The corresponding values can be found in table \ref{parameters}, with the particular feature that the effective total rest mass is close to $1.2 Gev$}
\label{sigma}
\end{figure}


\begin{center}

\begin{table}
\begin{center}
  \begin{tabular}{ccc}
$J^{P}$&$\omega$ (Mev$/\hbar$)& $M/1.2$ Gev\\
\\
$\frac{1}{2}^{+}$&96&1.00\\
$\frac{1}{2}^{-}$&184&1.00\\
$\frac{3}{2}^{+}$&187&1.27\\
$\frac{3}{2}^{-}$&179&0.93\\
$\frac{5}{2}^{+}$&137&1.11\\
$\frac{5}{2}^{-}$&137&1.03\\
\end{tabular}
\end{center}
\caption{Table of parameters}
\label{parameters}
\end{table}

\end{center}


\subsection{The two-particle Dirac oscillator}

Now we proceed to generalize the DMO to two particles. There is more than one generalization which gives a solvable two-particle problem, as can be seen in \cite{moshbook}, \cite{list}. Here we concentrate in one possibility for the interacting potential. For simplicity, let us set the masses of the particles as unity. It is also convenient to restore our frequency $\omega$ in this section, as we want to analyze the spectrum in terms of the coupling. After all, one can always go back to the former units by replacing $\omega \mapsto \omega \hbar / m c^2$. The hamiltonian and the Poincare invariant equation are, respectively

\bea
H= (\bfalpha_1 - \bfalpha_2) \cdot ( \v p - i \frac{\omega}{2} \v r B ) + \beta_1 + \beta_2
\label{eee19}
\eea
\bea
\left[ \sum_{s=1,2} \Gamma_s \left( \gamma_{s}^{\mu}(p_{\mu s} - i\omega x'_{\perp \mu s} \Gamma ) + 1 \right) \right] \Psi = 0
\label{eee20}
\eea
The interaction matrix $B$ is chosen here as $\beta_1 \beta_2 \gamma_{51}  \gamma_{52}$ with $\gamma_{51}= \gamma_5 \otimes \v 1$, similarly for $ \gamma_{52}$. In this case, the hamiltonian given above admits an expression which is quite similar to that of our previous algebraic analysis. With the appropriate definitions of $*$-spin operators $\Sigma^1_{\pm}, \Sigma^2_{\pm}, \Sigma^1_{3}, \Sigma^2_{3}$ for particles 1 and 2 and the bosonic operators $\v a, \v a^{\dagger}$ for the relative coordinate, one has 

\bea 
H= \Sigma^1_+ (\v s_1 \cdot \v a) + h.c. + \Sigma^1_3 - \Sigma^2_+ (\v s_2 \cdot \v a) - h.c. + \Sigma^2_3
\label{star2}
\eea
showing clearly that the structure leading to infinite degeneracy is still present through $\v s_i \cdot \v a$. The infinite degeneracy of the cockroach nest manifests itself by the "wrong" addition of energies. One could repeat the treatment given before in terms of invariants, which are still easy to identify. However, the original work of Moshinsky (described with detail in \cite{moshbook}) did not rely on this possibility and proceeded in the direction of decomposing the spinors and iterating the resulting equations connecting them (this corresponds, implicitly, to compute the fourth power of the hamiltonian). The resulting spectrum can be found as a function of the total angular momentum, total spin and total oscillator quanta: $ E= \pm E_{N,s,j,m}$ with

\bea
E_{N,s,j,m} =  \cases{ 2 \sqrt{1+\omega N} , 0 & for  $s=0, P=(-)^{j}$ \cr  2 \sqrt{1+\omega (N+2)}, 0 & for $s=1, P=(-)^{j}$ \cr 2 \sqrt{1+\omega (N+1)}, 0 & for $s=1, P=-(-)^{j}$}
\label{eee23}
\eea
The wavefunctions are known for all cases indicated before. Let $a_{\pm}, b_{\pm}, c_{\pm,\pm}$ be numerical coefficients; then we have the wavefunctions

\bea
\Psi= \left( \begin{array}{c} \psi_{11} \\ \psi_{21} \\ \psi_{12} \\ \psi_{22} \end{array} \right), \quad \left( \begin{array}{c} \psi_{11} \\ \psi_{22} \end{array} \right) = \frac{1}{\sqrt{2}} \left( \begin{array}{c} a_+ + a_- \\ a_+ - a_-  \end{array} \right) |N(j,0)jm \>
\label{eee24}
\eea
valid for $s=0$. Whenever $s=1$ and $P=(-)^j$, we have 

\bea
\left( \begin{array}{c} \psi_{11} \\ \psi_{22} \end{array} \right) = \frac{1}{\sqrt{2}} \left( \begin{array}{c} b_+ + b_- \\ b_+ - b_-  \end{array} \right) |N(j,1)jm \>
\label{eee26}
\eea
For $s=1$, $P=-(-)^j$ the result is

\bea
\nonumber \left( \begin{array}{c} \psi_{11} \\ \psi_{22} \end{array} \right) &=& \frac{1}{\sqrt{2}} \left( \begin{array}{c} c_{++} + c_{-+} \\ c_{-+} - c_{++}  \end{array} \right) |N(j+1,1)jm \> \\ &+& \frac{1}{\sqrt{2}} \left( \begin{array}{c} c_{+-} + c_{--} \\ c_{--} - c_{+-}  \end{array} \right) |N(j-1,1)jm \>
\label{eee27}
\eea
where the coefficients $c_{\pm \pm},a_{\pm},b_{\pm}$ are determined by the secular equations arising from
the Schroedinger equation for the relativistic hamiltonian. Taking into account (\ref{eee19}), the stationary equation
yields the complementary components of the wavefunction $\psi_{21}$,$\psi_{12}$. 
These wavefunctions can be found in terms of Racah coefficients in the appendix of \cite{phd}.

\subsection{The three-particle Dirac oscillator}

For this case, we follow again the book by Moshinsky \cite{moshbook} and recognize that the center of mass can be eliminated from the outset by proposing the following Poincare invariant equation and hamiltonian:

\bea
\left( n^{-1} \sum^{n}_{s=1} \Gamma_s(\gamma_s^{\mu} P_{\mu}) + \sum^{n}_{s=1} \left[ \gamma_s^{\mu} (p'_{\mu s}-i\omega x'_{\perp \mu s} \Gamma) +1 \right] \right) \Psi = 0
\label{ee31}
\eea

\bea
H \Psi = \sum_{s=1}^{n} \left[ \bfalpha_s \cdot (\v p'_s - i\omega \v x'_s B) + \beta_s \right]\Psi = E \Psi
\label{ee32}
\eea
where the primed observables denote the operators for a particle of index $s$ after subtracting the corresponding observable for the center of mass (either the momentum or the coordinate operators). Here, the matrix $B$ in the interaction is chosen as $\beta_1 \otimes ... \otimes \beta_n$, and we may choose in particular $n=3$. The spectrum is obtained by combining the equations for some of the spinor components of the wavefunction and by noting that the total number of quanta of two Fock states is conserved (corresponding to the oscillator states of the two remaining Jacobi coordinates). 
The wavefunctions are

\bea
\Psi_{+}=\left( \begin{array}{c} \psi_{111} \\ \psi_{122} \\ \psi_{212} \\ \psi_{221} \end{array}\right), \quad \Psi_{-}=\left( \begin{array}{c} \psi_{112} \\ \psi_{121} \\ \psi_{211} \\ \psi_{222} \end{array}\right)
\label{ee34}
\eea
and they satisfy

\bea
\ocal \Psi_+ = 0, \quad \ocal \equiv \v M D^{-1}_- \v M^{\dagger} - D_+ 
\label{ee37}
\eea
with 

\bea
\v D_+ = { \rm diag \ }(E-3,E+1,E+1,E+1),
\label{ee34bis}
\eea

\bea
\v D_+ = { \rm diag \ }(E-1,E-1,E-1,E+3)
\label{ee34bis}
\eea

\bea
\v M =2i \sqrt{2\omega} \left( \begin{array}{c c c c} \v S_3 \cdot \v a'_3 & \v S_2 \cdot \v a'_2 & \v S_1 \cdot \v a'_1 & 0 \\ \v S_2 \cdot \v a'_2 & \v S_3 \cdot \v a'_3 & 0 & \v S_1 \cdot \v a'_1 \\ \v S_1 \cdot \v a'_1 & 0 & \v S_3 \cdot \v a'_3 & \v S_2 \cdot \v a'_2  \\ 0 & \v S_1 \cdot \v a'_1 & \v S_2 \cdot \v a'_2 & \v S_3 \cdot \v a'_3  \end{array}\right),
\label{ee35}
\eea
with the operators without the center of mass defined as
\bea
\v a'_s & = & \v a_s - \frac{1}{3}(\v a_1 + \v a_2 + \v a_3).
\label{ee36}
\eea
In this treatment we recognize again the pattern of $*$-spin provided by the components $\Psi_{\pm} \propto | \pm \>$, leading to a hamiltonian of the form $H=\Sigma_{+} \v M + h.c. + \Sigma^1_3 +\Sigma^2_3+\Sigma^3_3$. Unfortunately, this does not lead to interpretations which are similar to our previous examples. The reason one has to diagonalize the operator $\ocal$ instead of using algebraic properties of the hamiltonian is related to integrability: In this case we have 8 invariant operators given by $P_{\mu},\v J, N$, while the total number of degrees of freedom (without taking into account the spin of the particles) is 9.

The application of this problem given by Moshinsky et al. \cite{moshbook} \cite{list} consisted on the calculation of the spectrum of masses of nucleons, together with a comparison with their experimental masses. There, the light quarks $u,d$ were treated as identical particles. Using the irreducible representations of the permutation group, suitable wavefunctions of the form Flavor $\times$ Spinor were used in the computation of energies and eigenfunctions. The application went as far as computing a form factor for the proton by using the information of the resulting wave function of the ground state. It is not our purpose to repeat such a prowess here. but instead let us compute the spectrum of this system without any other degrees of freedom than the ones provided from the outset. Our intention is to analyze the scaling properties when the coupling $\omega$ is varied from small to large values.
Using the states

\bea
\nonumber |n_1,l_1,n_2,l_2 (L); \frac{1}{2} \frac{1}{2} (T) \frac{1}{2} (S); J M \> = \qquad \qquad \\ \left[ \left[ (\dot r_1|n_1 l_1) \times (\dot r_2|n_2 l_2) \right]_L \times \left[ \left[ (1|\frac{1}{2})\times (2|\frac{1}{2}) \right]_T \times (3|\frac{1}{2})\right]_S \right]_{JM}
\label{ee38}
\eea

one can find the matrix elements of $\ocal$. The resulting
matrices are finite for each number of total quanta. We restrict to $N=0,1,2$. The wavefunctions can be finally obtained by finding the null vectors of
the matrix $\< \ocal \>$ for each energy. The complementary components are obtained as before, i.e. by using the original
stationary equation. See the table of states.

\begin{table}
\begin{tabular}[!h]{llllllllll}
  \hline  $N$ & \vline $N_1$ & \vline $N_2$ & \vline $n_1$ & \vline $n_2$ & \vline $l_1$ & \vline $l_2$ & \vline $P$ & \vline $L$ & \vline $J$ \\
  \hline  0 & \vline 0 & \vline 0 & \vline 0 & \vline 0 & \vline 0 & \vline 0 & \vline + & \vline 0 & \vline $S$  \\
  \hline  1 & \vline 1 & \vline 0 & \vline 0 & \vline 0 & \vline 1 & \vline 0 & \vline - & \vline 1 & \vline $|1-S| \leq J \leq 1+S $  \\
  \hline  1 & \vline 0 & \vline 1 & \vline 0 & \vline 0 & \vline 0 & \vline 1 & \vline - & \vline 1 & \vline $|1-S| \leq J \leq 1+S $  \\
  \hline  2 & \vline 2 & \vline 0 & \vline 1 & \vline 0 & \vline 0 & \vline 0 & \vline + & \vline 0 & \vline $S$ \\
  \hline  2 & \vline 0 & \vline 2 & \vline 0 & \vline 1 & \vline 0 & \vline 0 & \vline + & \vline 0 & \vline $S$ \\
  \hline  2 & \vline 1 & \vline 1 & \vline 0 & \vline 0 & \vline 1 & \vline 1 & \vline + & \vline $0 \leq L \leq 2 $ & \vline $|L-S| \leq J \leq L+S $ \\
  \hline  2 & \vline 2 & \vline 0 & \vline 0 & \vline 0 & \vline 2 & \vline 0 & \vline + & \vline 2 & \vline $|2-S| \leq J \leq 2+S $ \\
  \hline  2 & \vline 0 & \vline 2 & \vline 0 & \vline 0 & \vline 0 & \vline 2 & \vline + & \vline 2 & \vline $|2-S| \leq J \leq 2+S $ \\
 \hline
\end{tabular}
\caption{Table of states for $N_{max}=2$}
\label{states}
\end{table}

\begin{figure}[h]
\begin{center}
\includegraphics[width=6cm]{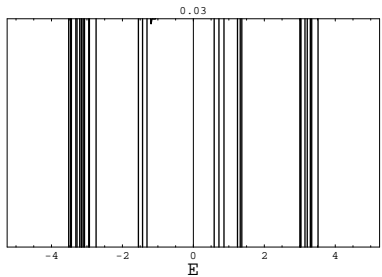}
\end{center}
\caption{Energies for $\omega=0.03$. The eigenvalues are distributed in four groups around the values $-3,-1,1,3$. The states coming from the cockroach nest lie around $1,-1$}
\label{spectrum31}
\end{figure}

\begin{figure}[h]
\begin{center}
\includegraphics[width=6cm]{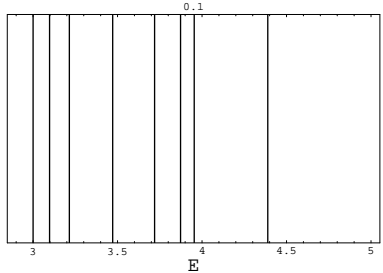}
\end{center}
\caption{Spectrum for $\omega=0.1$ and $N=2$. As the interaction increases, the levels become more spaced.}
\label{spectrum3part2}
\end{figure}

\begin{figure}[h]
\begin{center}
\includegraphics[width=6cm]{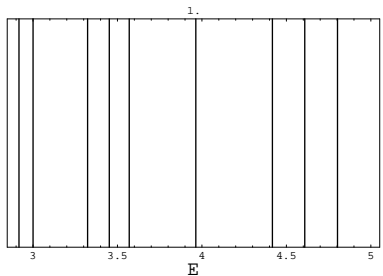}
\end{center}
\caption{Spectrum for $\omega=1$ and $N=2$. Here we see a typical spectrum for which the interaction does not allow a quasi relativistic expansion}
\label{spectrum3part3}
\end{figure}

In summary, the two and three particle Dirac oscillators can be solved. In the two-particle case, the features of the spectrum could be identified straightforwardly, given the simplicity of its hamiltonian. Moreover, we could show that such a hamiltonian obeys the algebraic scheme proposed for the one-particle case, exhaustively analyzed in section 1. The three-particle DMO constitutes a more challenging example, since the number of integrals of the motion does not match the total number of degrees of freedom. However, the numerical diagonalization can be done without much effort. The resulting spectra showed levels grouped around energies $1,-1$, which can be regarded as a consequence of the cockroach nest. In the following, we comment on the $n$ particle case, where we expect similar results.

\subsection{One dimensional $n$ particles}

Let us discuss briefly the infinite degeneracy present in this model. For our purposes, we may eliminate the spin of the particles by restricting ourselves to one-dimensional space. The only degrees of freedom remaining in our simplification are given in terms of the annihilation operators without center of mass and the $*$-spin operators associated to each particle. We argue that the kinetic part of the DMO hamiltonian

\bea
H= (\v 1 + B)\sum_{i}^{n} \sigma_{1}^{i} a'_{i} + h.c. + \rm{mass} 
\eea
is infinitely degenerate. To see this, we may apply eigenstates $| s^1_i \>$ of $\sigma_1^i$ to $(H-\rm{mass})^2$. Take states of the form $\psi= | s^1_1 \> ... | s^1_n \> \times | \dot N_{1} \> ... | \dot N_{n-1} \> $ where each $| \dot N_{i} \>$ is a Fock state with the only condition that $\sum \dot N_i = \mbox{constant}$. Clearly, the definition of our operators $a'_{i}$ gives $\sum a'_{i} = 0$ and any of the states proposed above for which all the $s^1_i$ are equal gives a vanishing kinetic term, regardless of the total number of quanta. The mass term thus not touch the oscillator parts, therefore the resulting matrix elements using these states lead to an infinitely degenerate spectrum. The cockroach nest makes itself present for an arbitrary number of interacting particles. Its
elimination is not a trivial task, in despite of our careful choice of observables. As a final question and in view of the results presented in this subsection, one may ask whether a system of one-dimensional Dirac particles can parallel the Calogero model \cite{calogero}, which is one of the most general integrable models in the non-relativistic realm.

To end this section let us quote Moshinsky regarding the applications of his work on relativistic oscillators:
"We conclude by stressing that we have made a calculation using a harmonic oscillator picture with a single parameter (frequency) and it is as good or as bad as many more complicated ones that start from QCD or that use many more parameters."

\section{Exactly solvable extensions}

In this section we present extensions of the one-particle DMO which allow solvability and connect our relativistic systems with implementations in Quantum Optics, via the Jaynes-Cummings model \cite{knight} for one and two atoms. The solvable extensions that we propose can be motivated entirely in the framework of non-local relativistic potentials \cite{yukawa1}, \cite{yukawa2}, \cite{nonlocalhiggs}, \cite{nonlocal1}, \cite{nonlocal2}. The key point in the introduction of such potentials is the presence of isospin: An internal degree of freedom of our fermion which we shall couple to an external field \cite{external}, \cite{conmau}. Interestingly, many of the problems related to infinite degeneracies of the DMO (discouraging the use of its wavefunctions as a basis for more complicated problems) can be removed in a natural way, preserving the simplicity of the model. We suggest the reader to follow references \cite{mau}.

Consider a hermitian operator of the form $\Phi(\v r,\v p)$ as the potential to be introduced in the total hamiltonian. One has $H^{(d)} = H^{(d)}_0 + \Phi$, with $H^{(d)}_0$ given by the $d-$dimensional Dirac-Moshinsky oscillator treated previously. On physical grounds, this corresponds to a bound fermion perturbed by a momentum-dependent potential. We introduce also an internal group for this field, for example the $SU(2)$ associated to isospin or as the gauge group of a non-abelian field. Let us denote the corresponding Pauli operators for isospin by $T_1, T_2, T_3$, with the usual definitions for the ladder operators $T_{\pm}$. The simplest expression that we can use is a linear one in $\v a$ and has the form

\bea
\Phi = \left( T_+ \v S \cdot \v a + T_- \v S \cdot \v a^{\dagger} + \gamma T_3 \right)
\label{01.2}
\eea
where we now use $\gamma$ to denote a coupling constant.
In fact, one may consider any potential of the form $\Phi=F(T_+ \v S \cdot \v a + T_- \v S \cdot \v a^{\dagger} + \gamma T_3)$ where $F$ is a function which admits a power expansion. Clearly, $[N + \ahalf T_3,\Phi]=0$ . A suitable group of states can be used to evaluate the $4 \times 4$ blocks of $H$. We describe this procedure by restricting ourselves to the linear case (\ref{01.2}) for simplicity. The lower dimensional examples follow the same pattern

\bea
H^{(1)}= \sigma_+ a + \sigma_- a^{\dagger} + m \sigma_3 + \nonumber \\
(A+\sigma_3 B) \left( T_+ a +T_ - a^{\dagger} + \gamma T_3 \right)
\label{01.41}
\eea
\bea
H^{(2)}= \sigma_+ a_r + \sigma_- a^{\dagger}_r + m \sigma_3 + \nonumber \\
(A+\sigma_3 B) \left( T_+ a_r + T_- a^{\dagger}_r + \gamma T_3 \right)
\label{01.42}
\eea
\bea
H^{(3)}= \Sigma_+ \v S \cdot \v a + \Sigma_- \v S \cdot \v a^{\dagger} + m \Sigma_3 + \nonumber \\
(A+\Sigma_3 B) \left( T_+ \v S \cdot \v a + T_- \v S \cdot \v a^{\dagger} + \gamma T_3 \right).
\label{01.43}
\eea
With these extensions, it is evident that the new invariants for one, two and three dimensions are

\bea
I^{(1)}= a^{\dagger} a + \frac{1}{2} \sigma_3 + \frac{1}{2} T_3
\label{01.5}
\eea

\bea
I^{(2)} = a_r a^{\dagger}_r + \frac{1}{2} \sigma_3, \quad J_3+\frac{1}{2} T_3 = a_r a^{\dagger}_r - a_l a^{\dagger}_l+ \frac{1}{2} \sigma_3 + \frac{1}{2} T_3
\label{01.6}
\eea

\bea
I^{(3)}= \v a^{\dagger} \cdot \v a + \frac{1}{2} \Sigma_3 + \frac{1}{2} T_3,
\quad \v J = \v a^{\dagger} \times \v a + \v S.
\label{01.7}
\eea

\subsection{Analytical Spectrum}

Now we compute the eigenstates of $H^{(3)}$. We evaluate the $4 \times 4$ matrix $H(N,j) \equiv \< \quad|H^{(3)}| \quad \>$.

\bea
|\phi^{N}_1\> = |n, (j+1/2,1/2) j, m_j\> |-\>_{\Sigma} |-\>_{T} \\ \nonumber
|\phi^{N}_2\> = |n, (j-1/2,1/2) j, m_j\> |-\>_{\Sigma}|+\>_{T} \\ \nonumber
|\phi^{N}_3\> = |n-1,(j-1/2,1/2) j,m_j\> |+\>_{\Sigma} |-\>_{T} \\ \nonumber
|\phi^{N}_4\> = |n-1,(j+1/2,1/2) j, m_j\> |+\>_{\Sigma}|+\>_{T}
\label{02.1}
\eea
where $n$ is the oscillator radial number, $j$ is the total angular momentum and $m_j$ its projection in the $z$ axis. These are eigenstates of $I^{(3)}$ with eigenvalue $N=2n+j-1/2$. 

The resulting $4\times 4$ blocks of $H$ with elements $H(N,j)_{kl}=\<\phi^{N}_k|H|\phi^{N}_l\>$ are

\bea
\left( \begin{array}{cccc} -m-(A-B)\gamma & (A-B)\sqrt{2(n+j)} & -\sqrt{2(n+j)} & 0 \\ (A-B)\sqrt{2(n+j)} & -m+(A-B)\gamma & 0 & \sqrt{2n} \\ -\sqrt{2(n+j)} & 0 & m-(A+B)\gamma & (A+B)\sqrt{2n} \\ 0 & \sqrt{2n} & (A+B)\sqrt{2n} & m+(A+B)\gamma \end{array} \right) \nonumber
\label{02.2}
\eea
and the energies can be obtained explicitly for each of these blocks using the formula for the roots of a quartic polynomial. The infinite degeneracy is now broken, since one cannot reduce $H(N)$ to smaller blocks where only $n$ appears.
The exception to this occurs when $A=B=0$, recovering the DMO without additional external fields.

\subsection{Lorentz invariant form and Pauli coupling revisited}

With the aid of a vector $u_{\mu}$ we can introduce more interactions in a covariant way. A non-local,
non-abelian field tensor $\fcal^{\mu \nu}=\sum_{i=1}^{3} T_i \fcal^{\mu \nu}_i$ can be introduced
in the equation by means of the Pauli coupling. We propose

\bea
\fcal_1^{\mu \nu} = \epsilon^{\mu \nu \lambda \rho} u_{\lambda} r_{\perp \rho} \\
\fcal_2^{\mu \nu} = \epsilon^{\mu \nu \lambda \rho} u_{\lambda} p_{\perp \rho} \\
\fcal_3^{\mu \nu} = 0,
\label{dos}
\eea
for which the Dirac equation reads
\bea
[\gamma_{\mu}p^{\mu} + m + S_{\mu \nu}F^{\mu \nu} + B S_{\mu \nu}\fcal^{\mu \nu}] \psi = 0,
\label{tres.1}
\eea

\begin{figure}[h]
\begin{center}
\includegraphics[scale=0.45]{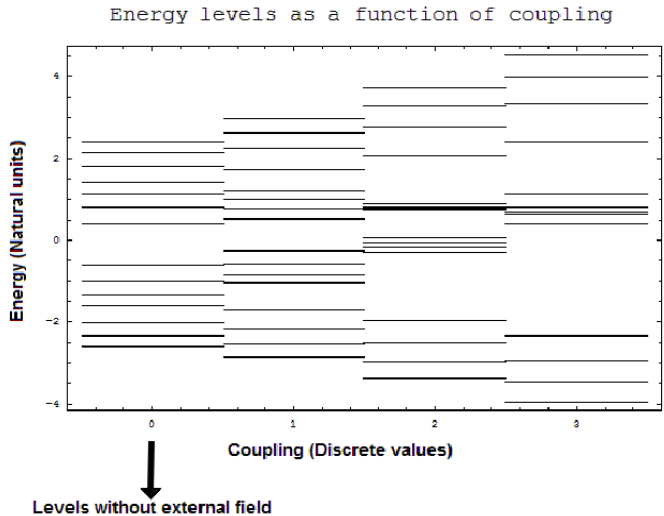}
\end{center}
\caption{Spectrum of our exactly solvable extension. The vanishing coupling shows the eigenvalues of the Dirac oscillator.
Degeneracies are lifted and level spacing increases.}
\label{perturbedeigenvalues}
\end{figure}

This type of fields have been introduced with the purpose of describing a finite characteristic length (due to non-locality) and also as a way to prevent divergences in perturbation theory.
The nature of such fields can be elucidated by inserting our $\fcal_{\mu \nu}$
in the corresponding non-local field equations \cite{nonlocal1}, \cite{nonlocal2}, \cite{nonlocal3}. Using 

\bea
\fcal^{\mu \nu}= u^{\mu}(r^{\nu}_{\perp}T_1+p^{\nu}_{\perp}T_2) - \mu \leftrightarrow \nu 
\eea
one has

\bea
\fcal^{\mu \nu} = i([p^{\mu},B^{\nu}]-\mu \leftrightarrow \nu) + [B_{\mu},B_{\nu}].
\label{cinco.1}
\eea
The gauge potential and the current can be obtained in the form

\bea
B_{\mu}=u_{\mu} (\frac{1}{2}  r_{\mu} r^{\nu}_{\perp} T_1 + r_{\nu} p^{\nu}_{\perp} T_2 ) \qquad \mbox{Bilinear in $p$, $r$}.
\label{cinco.2}
\eea
\bea
j^{\nu} = i[p_{\mu},\tilde{\fcal}^{\mu \nu}] + [B_{\mu},\tilde{\fcal}^{\mu \nu}]
\eea

\bea
=-u^{\nu} T_1 + p^{\nu}_{\perp} + \left( \frac{1}{2} \lbrace p_{\perp}^{\nu}, r_{\mu} r^{\mu}_{\perp} \rbrace - \lbrace p_{\perp}^{\mu}, r_{\perp}^{\nu} \rbrace r_{\mu} \right) T_2 \\ \nonumber
=-u^{\nu} T_1 + p^{\nu}_{\perp} + \mbox{ trilinear terms in $p$,$r$ }.
\label{cinco.3}
\eea
We do not elaborate further on these points, since our aim here is to simply note that our momentum-dependent potentials admit a treatment which is parallel to that given in section 1. We refer the reader to \cite{conmau} for detailed derivations. The eigenvalues for the one-dimensional extension are given in the figure \ref{perturbedeigenvalues}.

In this section we have shown that the introduction of extra degrees of freedom motivated by particle physics (e.g. isospin) can be used to propose exactly solvable models. Furthermore, such extensions can be rewritten in Lorentz invariant form and the coupling of the external fields to our DMO was shown to parallel the Pauli coupling treated in section 1. Also, the removal of infinite degeneracy was possible by extending the space (rather than introducing a $j$ dependence in the interactions). Although we have computed the spectrum analytically, it is desirable to understand the dynamics of this system. Such a problem can be analyzed by methods dealing with entanglement of the different observables of our system: spin, $*$-spin, isospin and oscillator operators.

\subsection{Quantum Optics}

The remarkable analogy
between the Dirac oscillator and the Jaynes-Cummings hamiltonian has been
pointed out before \cite{solano}, \cite{delgado}, \cite{delgado2} with the aim of producing such a system in a quantum
optical experiment.
The structure of our extended hamiltonian shows that our model 
can be mapped to a Jaynes-Cummings hamiltonian of two atoms (of two levels each) inside an electromagnetic cavity. If the dimensions of such a cavity are properly chosen, the eigenmodes of the quantized electromagnatic field will be sufficiently separated in frequency, with the possibility of coupling our atoms to only one boson operator. The one-dimensional example and the double Jaynes-Cummings model coincide: 

\bea
H= \sigma_+ a + \sigma_- a^{\dagger} + m \sigma_3 + T_+ a +T_- a^{\dagger} + \gamma T_3
\label{01.41}
\eea
where we have to identify $\sigma,T$ with the operators for the two-level atoms 1 and 2. The operator $a$ is now the annihilation operator of the electromagnetic field mode. Spin-spin interactions can be introduced as well.

\subsubsection{Dynamical application: Entanglement and Decoherence}

The origin of entanglement and decoherence measures is related to quantum information and quantum computation \cite{quang}. However, such quantities can be defined and computed in such a simple way that they can be used to analyze the dynamical features of general systems involving several degrees of freedom. Here we shall take advantage of this situation and proceed to
define a partition of the system $A+B$.

We take a pure state density operator 
$\rho = | \psi(t) \rangle \langle \psi(t) |$ of the entire system
and compute  purity $P$ and entropy $S$ of the Dirac oscillator subsystem.

\bea
P(t)&=& \rm{Tr}_{N,\sigma} \left( \left( \rm{Tr}_{\tau} \rho(t) \right)^2 \right)\nonumber \\
S(t)&=& - \rm{Tr}_{N,\sigma} \left( \rm{Tr}_{\tau} \rho(t) \rm{Log}\left( \rm{Tr}_{\tau} \rho(t) \right) \right),\nonumber \\
\label{4.3}
\eea
where $\rm{Tr}_{N,\sigma}$ is the trace with respect to oscillator and 
$*$-spin degrees of freedom, while $\rm{Tr}_{\tau}$ is the trace 
with respect to isospin.
Let us analyze the one dimensional case for simplicity. The integral of the motion is

\bea
I^{(1)}= a^{\dagger} a + \frac{1}{2} \sigma_3 + \frac{1}{2} T_3
\label{01.5}
\eea
We use the eigenstates of $I^{(1)}$
\bea
|\phi_1^n\> = |n+2\> |--\>
\quad\quad
|\phi_2^n\> = |n+1\> |-+\> \nonumber\\
|\phi_3^n\> = |n+1\> |+-\>
\quad\quad
|\phi_4^n\> = |n\> |++\>
\label{8}
\eea
\bea
H=\left(
\begin{array}{cccc}
H_0&0&0&\dots\\
0&H_1&0&\dots\\
0&0&H_2&\\
\vdots&\vdots&&\ddots
\end{array}
\right),
\label{10}
\eea
where $H_n$ is a $4 \times 4$ block.

Now we analyze the entanglement of a Dirac oscillator with the external field. the initial state is chosen as $\psi=\chi_n 
\otimes \chi$, 

\bea
|\chi\>=1/\sqrt{2}(\cos{\theta}|+\>+\sin{\theta}|-\>)
\eea
and $\chi_n$ is a solution of the unperturbed Dirac oscillator

\bea
|\chi_n\> = 
A_n^{(+)} |n\>|+\> + A_n^{(-)} |n+1\>|-\>
\label{4.1}
\eea
We use the exact energies and wavefunctions to compute $P(t), S(t)$ (purity and entropy). Other initial conditions
can be used in the context of Quantum Optics, for example in cases where the initial state is prepared as a product of the two atoms $| \pm \>| \pm \> $. In that case, the external field induces entanglement between such degrees of freedom, although the atoms do not interact directly but only through the cavity. However, this side of the analogy will not be discussed here \cite{mau} \cite{conmau}.
here.

\begin{figure}[h]
\begin{center}
\begin{tabular}{ll}
\includegraphics[scale=.42]{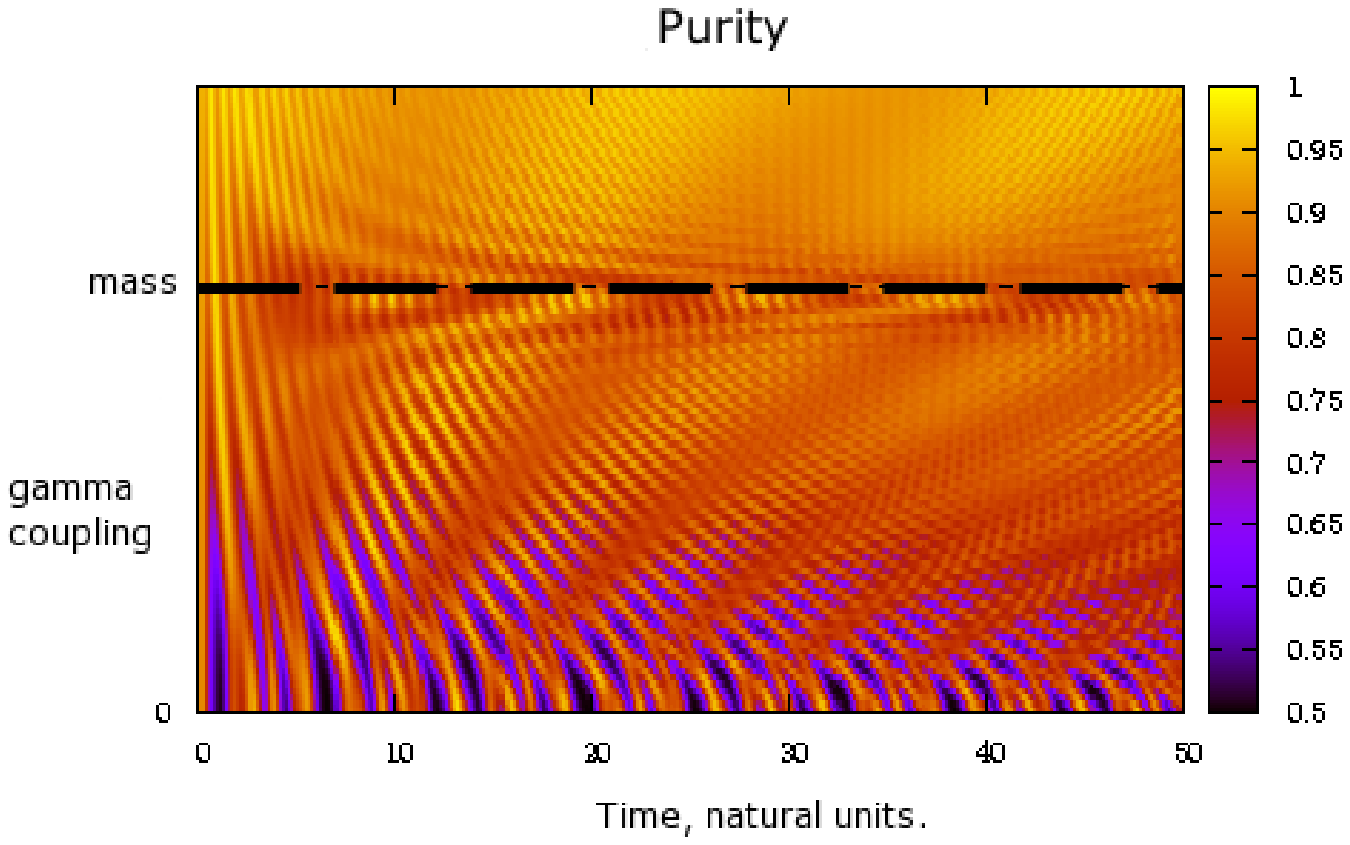} &
\includegraphics[scale=.42]{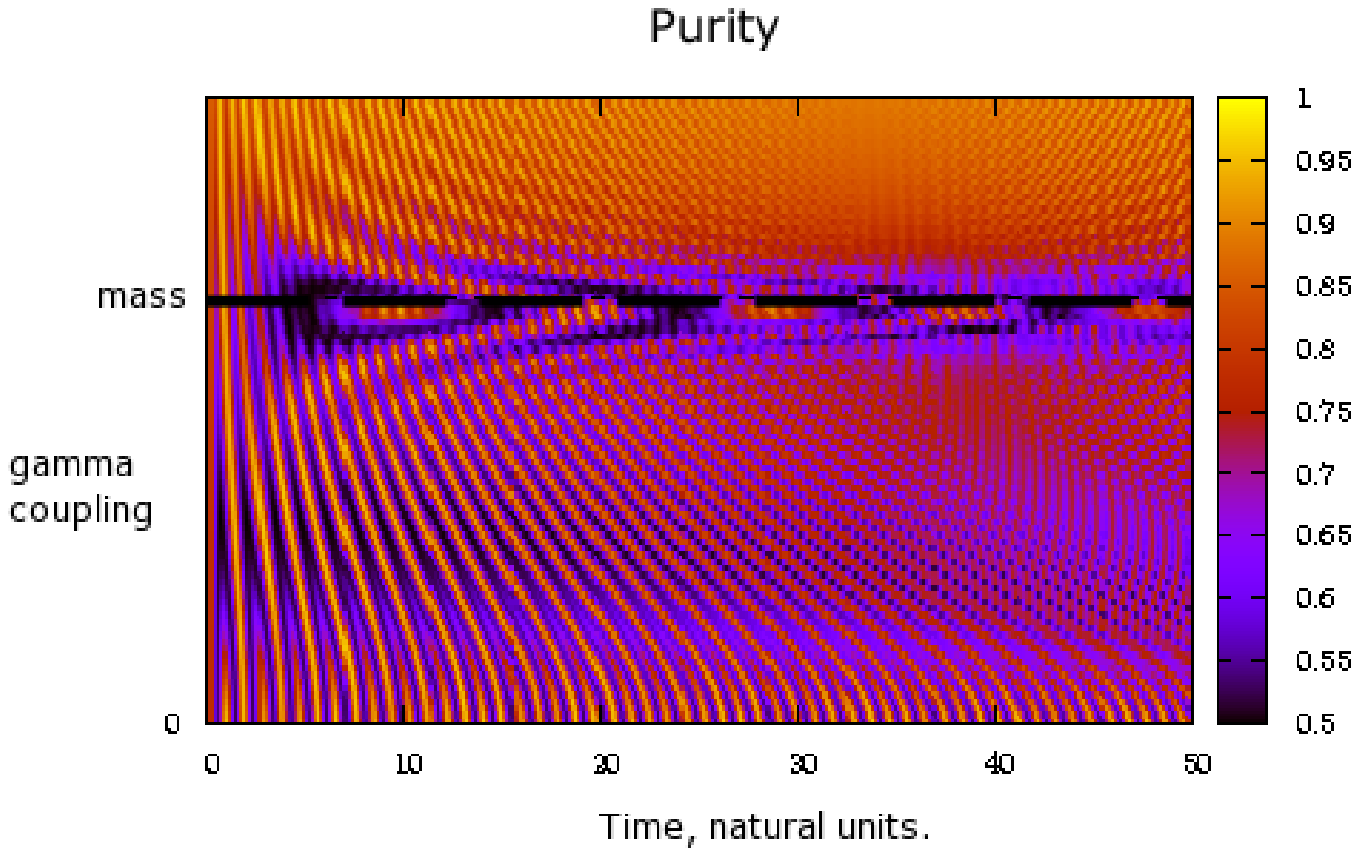}
\end{tabular}
\end{center}
\caption{A resonant effect around $\gamma=m$. The field produces entanglement in a regime where the energy is nearly the rest mass. We have used purity as the simplest way to characterize the entangled state.}
\label{purity}
\end{figure}

\begin{figure}[h]
\begin{center}
\begin{tabular}{ll}
\includegraphics[scale=.42]{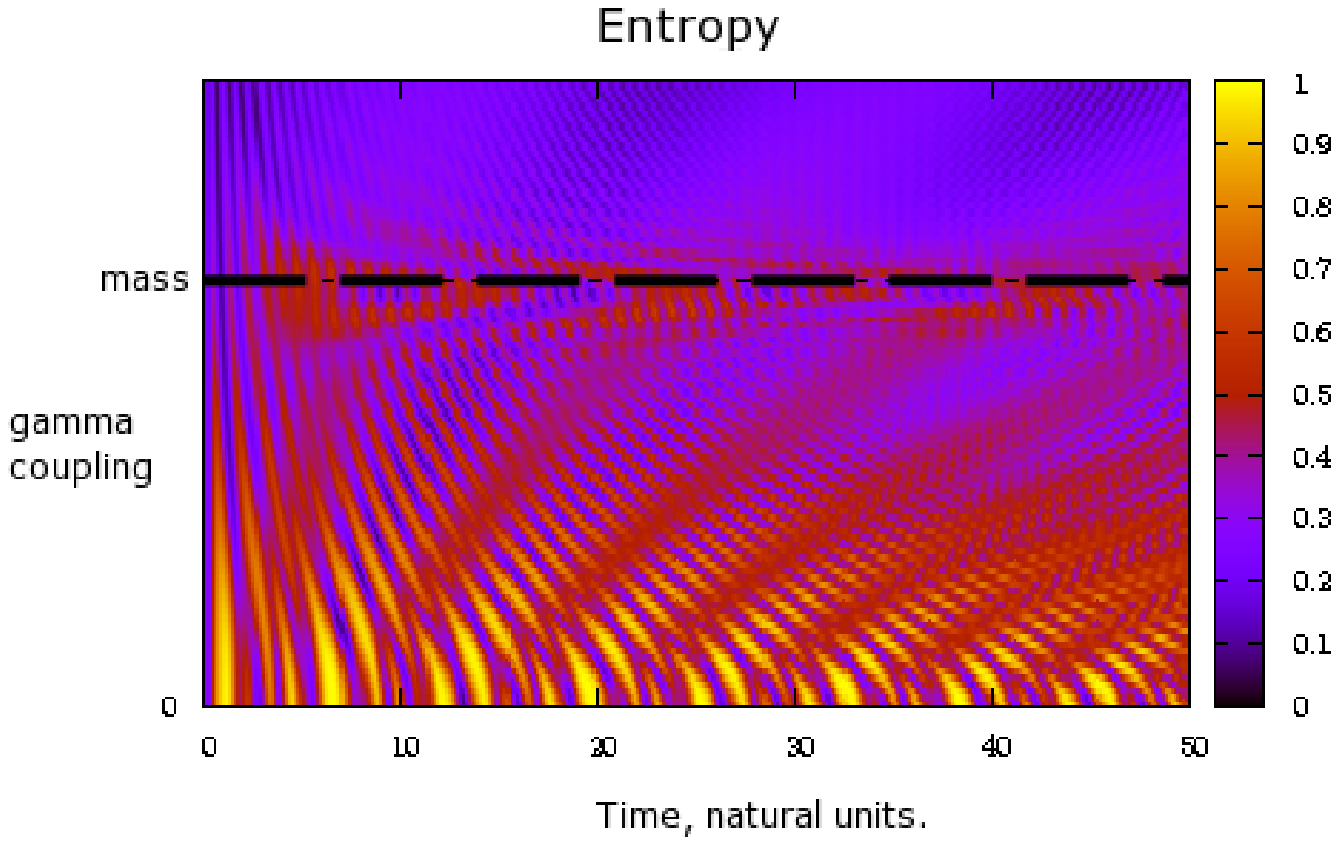} &
\includegraphics[scale=.42]{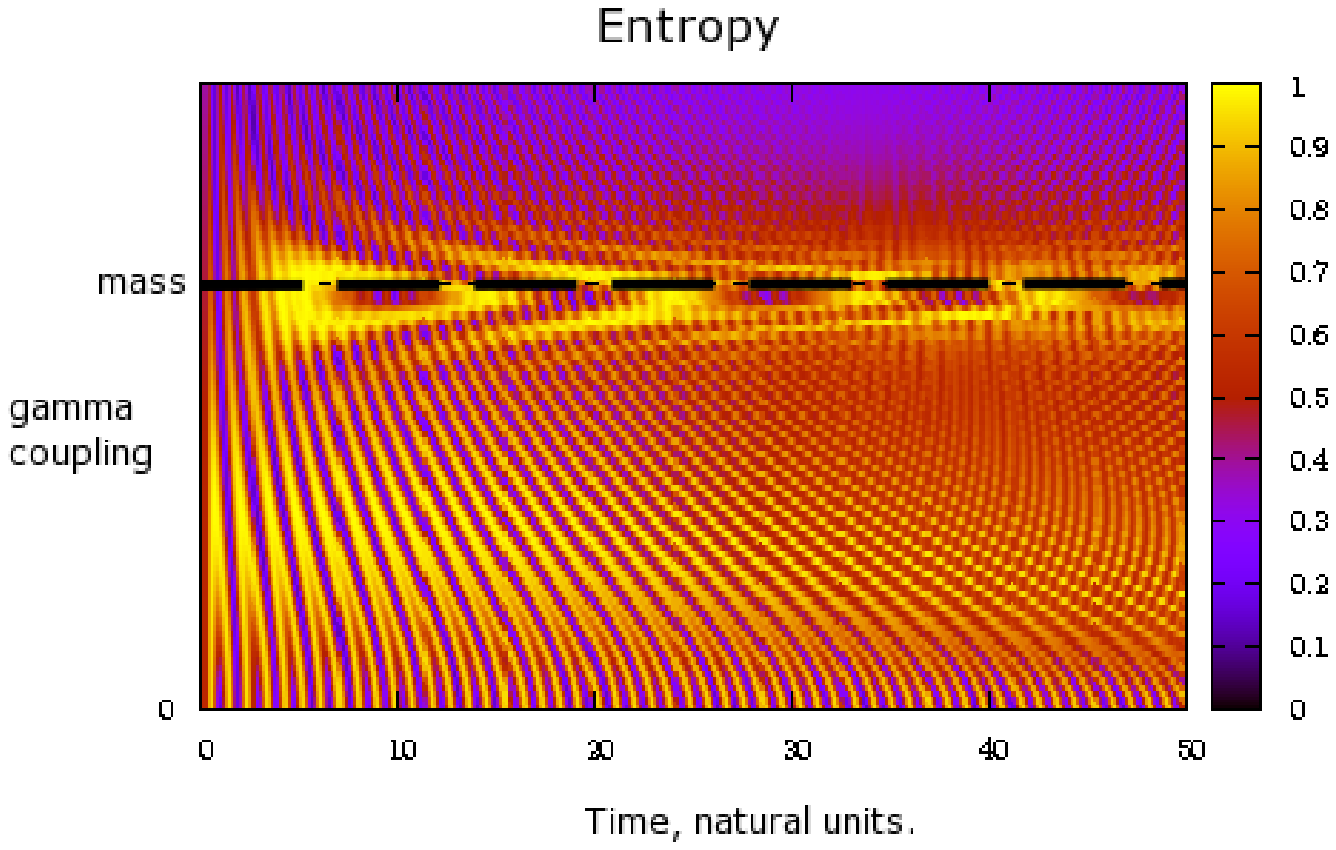}
\end{tabular}
\end{center}
\caption{A resonant effect around $\gamma=m$. The field produces entanglement in a regime where the energy is nearly the rest mass. Here we show the entropy by way of comparison. The structure obtained in figure \ref{purity} is also present in this result}
\label{entropy}
\end{figure}

We have learned in this section that the path of algebraic solvability leads to many possibilities regarding relativistic wave equations with additional degrees of freedom such as isopin and their ad hoc realization in quantum optical experiments where atomic levels can be used to emulate certain observables. 

In our dynamical study, we have seen that the perturbation of a DMO with fields of arbitrary intensity admits an analytical treatment, with the conclusion that the rest mass of the system responds to the stimulus via entanglement. Our toy model suggests that particle creation and maximal entanglement are related. On the other hand, this can be interpreted as a resonant effect in the Quantum Optics analogy.

In what touches the experimental realization of this model, the following setup can be proposed with quite general parameters. We require two atoms of different species trapped in an
electromagnetic cavity. The coupling constant $\gamma$ defined in relation with $m$ (fixed as unity), can be
adjusted by placing the atoms in regions with different field intensities. Moreover, a large
distance between the atoms is needed in order to ignore direct interaction terms. The
evolution of entanglement studied in this section can be realized by preparing the initial
state. In principle, one could prepare it by trapping one atom in the cavity
and measuring the energy of the total system (this corresponds to a dressed state of the
Jaynes-Cummings model with one atom). After this is achieved, the second atom can be
introduced in the trap. This setup thus emulates our model and makes it experimentally
accessible.

 \section[Emulating a DMO in Electromagnetic Billiards]{Emulating a Dirac-Moshinsky oscillator in Electromagnetic Billiards}

In this section we give an account of our recent findings related to hexagonal lattices and the emulation of Dirac equations. In recent years, there has been an explosion of papers (for instance, \cite{novoselov} and references cited therein) related to the experimental observation of true monolayers of carbon obtained from graphite: graphene. The technique, known as micromechanical cleavage, takes advantage of the property that graphite is composed of weakly interacting layers of carbon atoms and such layers can be removed and analyzed individually in atomic microscopes (graphene flakes). Together with the many possibilities for the practical applications of such materials, there is an additional feature which is of particular interest to our subject: Relativistic quantum mechanics. 

The band theory of graphite was studied by Wallace in the 1950's \cite{wallace}. There it was shown that the dispersion relation of electrons propagating in a hexagonal lattice becomes linear at the edges of the Brillouin zone - a hint for a relativistic energy formula, although the slope is given by the Fermi velocity instead of the speed of light. At the corners of the Brillouin zone, one could find conical energy surfaces and, according to the quantum field theoretical approach proposed by Semenoff \cite{semenoff}, one could also find a pseudospin from the decomposition of the hexagonal lattice into two triangular sublattices. Since it is the hexagonal structure what is essential to this analogy, one could also try to emulate the same behavior by propagating waves in periodic arrays of resonators - whose resonances should play the role of the atomic orbitals in a material. This happy analogy meets the existing technology of microwave cavities, originally used in the context of chaotic billiards. Here we review the corresponding theoretical treatment in a detailed manner and go further by proposing a model of deformations giving rise to an effective wave equation given by the Dirac-Moshinsky oscillator. 

\begin{figure}[h]
\begin{center}
\includegraphics[scale=0.5]{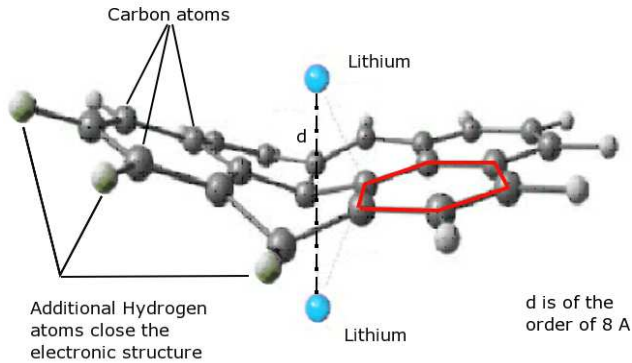}
\end{center}
\caption{Deformation of a graphene sheet, induced by the proximity of Lithium atoms (blue spheres). A hexagonal cell is shown in red. This plot can be obtained using Density Functional calculations to determine the equilibrium positions of the atomic centers. See, for example, Seligman and Jalbout \cite{dft}).}
\label{jalbout}
\end{figure}

 \subsection{One-dimensional Dirac equation}

 The situation described above can be modelled in a simple way by a Schroedinger equation with
 a potential consisting of deep wells, each of them located at a lattice point.
The specific shape of atomic wave functions is irrelevant, as long as we know how the
overlaps (interactions) decay as a function of the distance between resonators.
For practical purposes, such decay can be regarded as exponential, which follows
from considering a lattice of constant potential wells. As an additional remark,
such potentials should be deep enough such that only one level (or isolated resonance)
contributes to the dynamics.

A lattice consisting of two periodic sublattices is considered.
They have the same period and are denoted as type A and type B.
Each sublattice point can be labeled by an integer $n$ according to its position
on the line, \ie $x_n$. The energy of the single level to be considered
in the well is denoted by $\alpha$ for type A and $\beta$ for type B.
The state corresponding to a particle in site $n$ of lattice A is denoted by $|n\>_A$
and the corresponding localized wave function is given by $\xi_A(x-x_n) = \<x|n\>_A$.
The same applies for B. The probability amplitude $\Delta$ (or overlap) between nearest
neighbors is taken as a real constant.

\bea
H= \left( \begin{array}{cc} H_{AA}& H_{AB} \\ H_{BA}& H_{BB}\end{array} \right)
\label{1}
\eea

\begin{figure}[h]
\begin{center}
\includegraphics[scale=0.6]{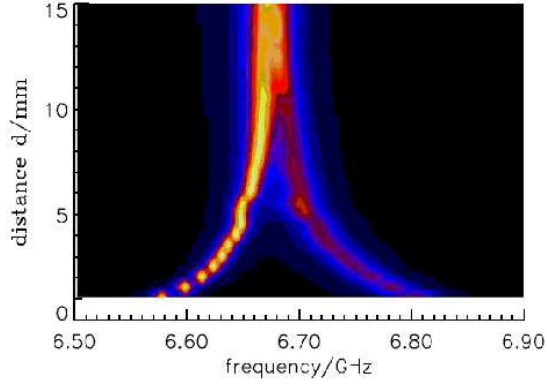}
\end{center}
\caption{Density plot of the coupling between two resonators. The exponential decay of the coupling is demonstrated by noting that two resonators constitute a two level system for which $E=E_0 \pm \Delta$. As the distance increases, the level splitting $\Delta$ tends to zero and the two peaks (orange paths) merge into a single peak exponentially fast. Courtesy of U. Kuhl.}
\label{expdensity}
\end{figure}

\begin{figure}[h]
\begin{center}
\includegraphics[scale=0.25]{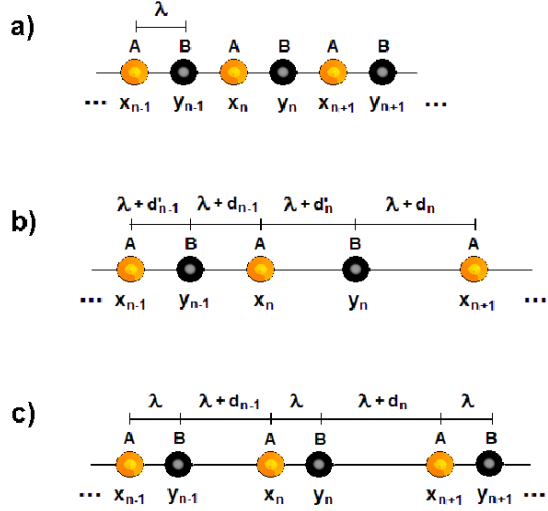}
\end{center}
\caption{Configuration of potential wells (or resonators) on a chain. We use yellow and black discs to recall that in principle the resonators can be of different type. a) The periodic case. b) General deformation. c) Dimer deformation.}
\label{chain}
\end{figure}

\begin{figure}[h]
\begin{center}
\includegraphics[scale=0.6]{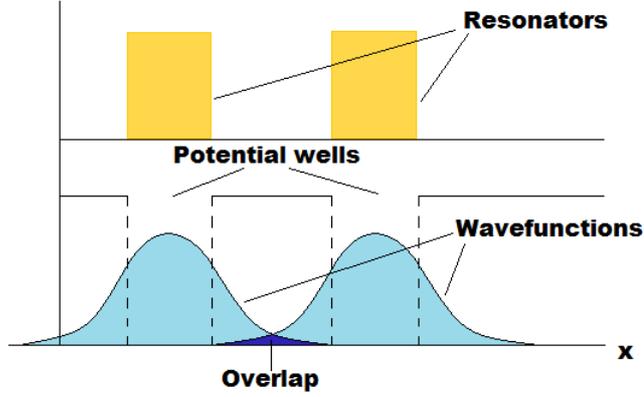}
\end{center}
\caption{Resonators in a one dimensional lattice. The plot above gives a representation of
resonators as a function of the x-coordinate, while the plot below shows an
idealization of the corresponding potential (wells) and the wave functions of
resonances. These functions may leak outside the wells.}
\label{resonators}
\end{figure}

The hamiltonian of a tight-binding chain can be cast in terms of Pauli matrices
$\sigma_3,\sigma_+ =\sigma_1+i\sigma_2, \sigma_-=\sigma_+^{\dagger}$ by defining

\bea
\Pi = \left( \begin{array}{cccc} \ddots & & & \\ & \Delta & \Delta & \\ &&\Delta&\Delta \\ &&& \ddots \end{array} \right)
\label{3}
\eea
and setting $M=(\alpha-\beta)/2, E_0=(\alpha+\beta)/2$. We have 

\bea
H= E_0 + \sigma_3 M + \sigma_+ \Pi + \sigma_- \Pi^{\dagger}
\label{4}
\eea
This is a general structure which explains the appearance of pseudospin.

It is left to show that there is a region where the spectrum is linear (Dirac). 
The spectrum is computed by squaring $H$. 
\bea
(H-E_0)^2 = M^2 + \Pi \Pi^{\dagger}
\label{6}
\eea
Bloch's theorem enters in the form

\bea
\Pi \phi_k = \Delta (1+e^{i2\pi \lambda k}) \phi_k, \quad \Pi \Pi^{\dagger} \phi_k = \Delta^2 |1+e^{i2\pi \lambda k}|^2 \phi_k
\label{7}
\eea
The energies and eigenfunctions of $H$ are

\bea
E(k) = E_0 \pm \sqrt{\Delta^2|1+e^{i2\pi \lambda k}|^2 + M^2} 
\label{9}
\eea
\bea
\psi^{\pm} = N \left(\begin{array}{c} \phi_k \\ \frac{\pm E(k) - E_0 - M}{\Delta(1+e^{i2\pi \lambda k})} \phi_k \end{array} \right),
\eea
Around points where the inter-band distance is minimal, we have the usual relativistic formula

\bea
E(\kappa)= E_0 \pm \sqrt{\Delta^2 \kappa^2 + M^2},
\label{10}
\eea

The amplitudes are proportional to the overlap between neighboring sites and decay
exponentially as a function of the separation distance between resonators, \ie

\bea
\Delta_{n,n+1} = \Delta e^{-d_n / \Lambda},
\label{12}
\eea
where $d_n + \lambda$ is the separation distance between resonators of type A and B
in the $n$-th position. When $d_n=0$, the periodic configuration is recovered.
The length $\Lambda$ has been introduced for phenomenological reasons: The decay
law might be given by a multipole law, but we fit it to an exponential decay by
adjusting $\Lambda$.

With all this, it is natural to expect a modification in the operators $\Pi,\Pi^{\dagger}$.
We use $a, a^{\dagger}$ and impose $[a, a^{\dagger}]=\omega = constant$
(The limit $\omega=0$ recovers Bloch's theorem). One finds the conditions

\bea
\Delta_{n,n} = \Delta,  \qquad \Delta^2_{n+1,n+2}- \Delta^2_{n,n+1}= \omega
\label{15}
\eea
Therefore the distance formula for the resonators is

\bea
d_n = \Lambda \log{\left( \frac{\Delta^2}{\Delta^2-n \omega} \right)}, \qquad  0<n<n_{max}
\label{21}
\eea
with $n_{max}= [|\frac{\Delta^2}{\omega}|]$.

Finally, we have the hamiltonian

\bea
H= E_0 + \sigma_3 M + \sigma_+ a + \sigma_- a^{\dagger}
\label{22}
\eea

with energies and wave functions

\bea
E(n) = E_0 \pm \sqrt{\omega n + M^2}, \qquad  0 > n > \Delta^2 / \omega \\ \nonumber
\psi^{\pm} = N \left(\begin{array}{c} \phi_{n+1} \\ \frac{\pm (E(n) - E_0) - M}{\sqrt{\omega (n+1)}} \phi_n \end{array} \right),
\label{24}
\eea

\begin{figure}
\begin{center} 
\includegraphics[scale=0.8]{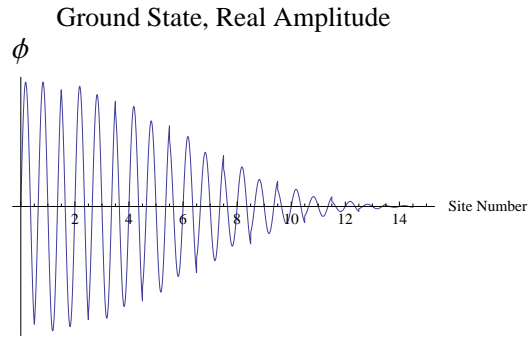}
\end{center}
\caption{Ground state as a function of site number. The ground state wavefunction
is obtained by multiplying the values given in the ordinate by the individual
resonant wavefunctions. These are considered to be highly peaked at each site.
The signs alternate from site to site. The envelope is approximately gaussian (nodes are absent).}
\label{groundstate11}
\end{figure}

\begin{figure}
\begin{center}
\includegraphics[scale=0.8]{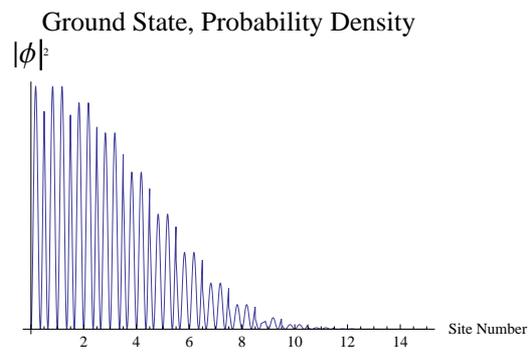}
\end{center}
\caption{Ground state density as a function of site number. The probability density
is obtained by multiplying the values in the ordinate by the individual resonant
wavefunctions, which are considered to be highly peaked at each site.
The density has a gaussian envelope and does not exhibit nodes.}
\label{groundstate12}
\end{figure}

\subsection{Two-dimensional Dirac equation}


The concepts given in the last section are now extended to produce an emulation
of graphene. We shall use the same algebraic strategy to derive spectra and a
possible extension through deformations, namely the two dimensional Dirac-Moshinsky oscillator.

\begin{figure}[h]
\begin{center}
\includegraphics[scale=0.2]{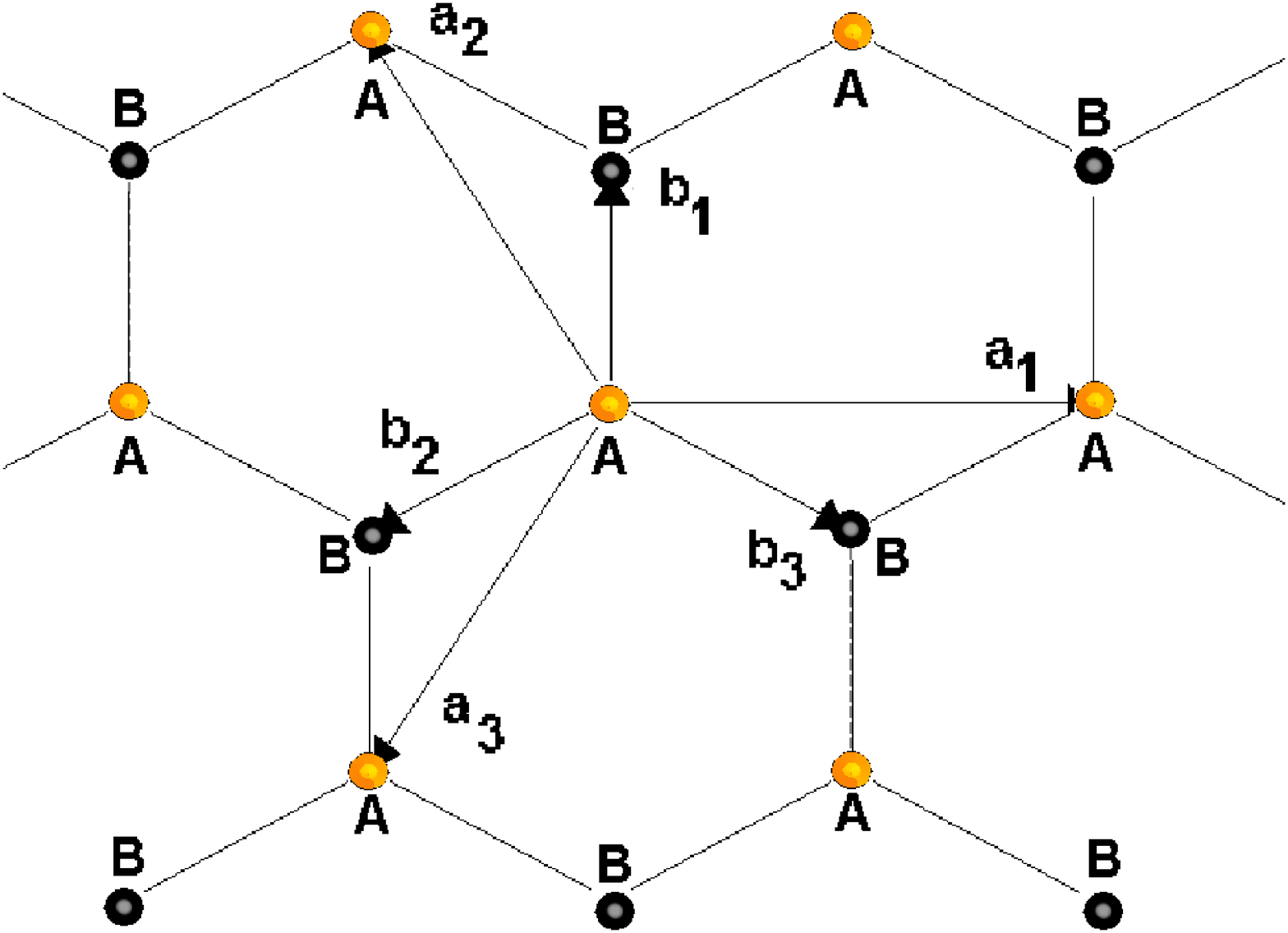}
\end{center}
\caption{Vectors describing a 2D array. The components of $\v b_i$ and $\v a_i$ are given in the text.}
\label{fig2d}
\end{figure}

\begin{figure}[h]
\begin{center}
\includegraphics[scale=0.25]{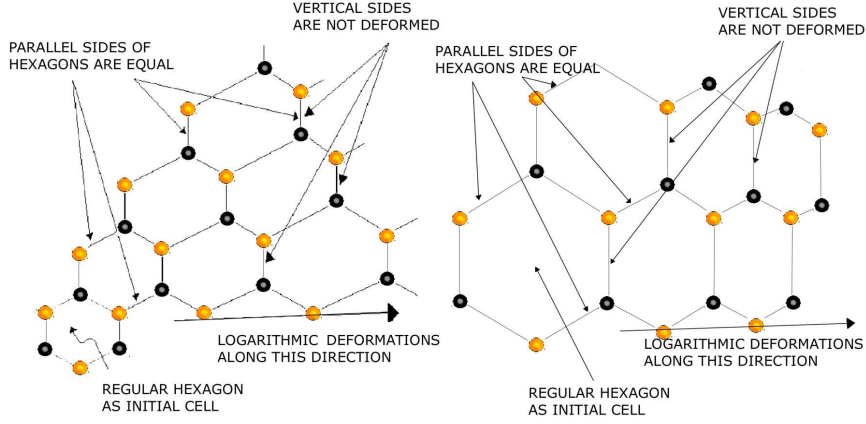}
\end{center}
\caption{Two possible deformations of the lattice}
\label{mesh2}
\end{figure}

\begin{figure}[h]
\begin{center}
\includegraphics[scale=0.4]{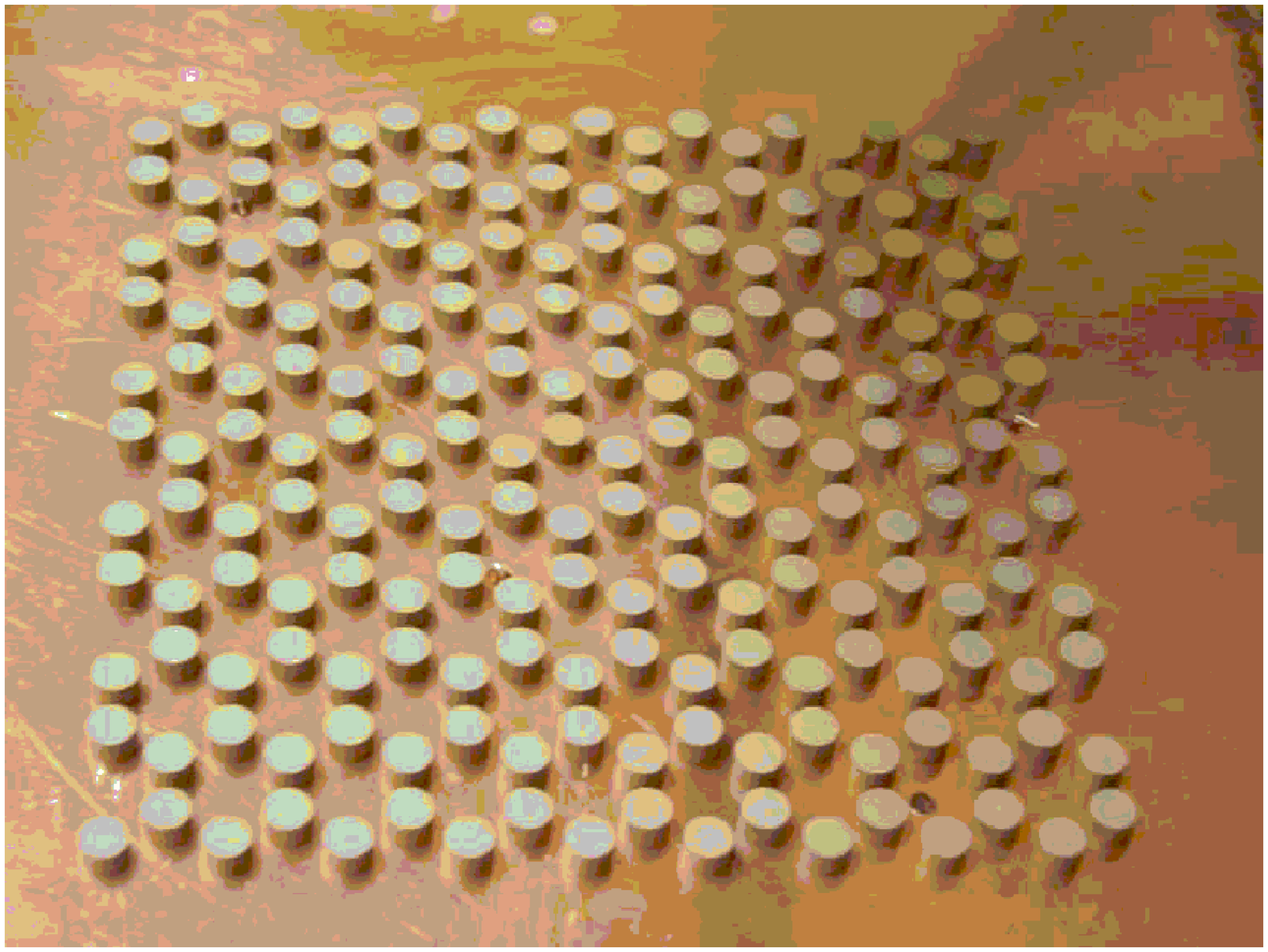}
\end{center}
\caption{Resonators with high dieletric constant $\epsilon \sim 34$. Courtesy of F. Mortessagne}
\label{resonators}
\end{figure}

\subsubsection{The free case in 2D}

We start with the definition of the vectors which generate our hexagonal lattice (see figure \ref{fig2d}). It is divided in two triangular sublattices, one of them generated by $\v a_1=(\sqrt{3},0),\v a_2=(-\sqrt{3}/2,3/2), \v a_3=(-\sqrt{3}/2,-3/2)$ (grid A) while the other sublattice is obtained by adding the vectors $\v b_1=(0,1), \v b_2=(-\sqrt{3}/2,-1/2), \v b_3=(\sqrt{3}/2,-1/2)$. These vectors are so far dimensionless. The position vectors $\v r_A, \v r_B$ of the periodic lattices are obtained by introducing the factor $\lambda$ (with the dimensions of length). Deformed lattices can also be described by these vectors, but the position vectors become more complicated functions of $\v a_i, \v b_i$. We denote by $\v A$ the vector parametrizing sublattice A. For B we use $\v A + \v b_1$. The state vectors (eigenvectors) for individual potential wells on grid A shall be denoted by $| \v A \>$, giving wave functions of individual wells as $\xi_A(\v r - \v r_A)=\< \v r | \v A \>$. For grid B we use $| \v A + \v b_1\>$. The tight binding hamiltonian in this case is given by

\bea
H &=& \alpha \sum_{\v A} |\v A \> \< \v A | + \beta \sum_{\v A} |\v A + \v b_1 \> \< \v A + \v b_1 |  + \nonumber \\ 
&+& \sum_{\v A, i=1,2,3} \Delta \left( |\v A \> \< \v A + \v b_i | + |\v A + \v b_i \> \< \v A | \right)
\label{28}
\eea
The usual Pauli operators are constructed through the definitions

\bea
\sigma_{+} = \sum_{\v A} |\v A \> \< \v A + \v b_1 |, \qquad \sigma_{-} = \sigma_{+}^{\dagger} \\ \nonumber
\eea

\bea
\sigma_3 = \sum_{\v A} |\v A \> \< \v A | - |\v A + \v b_1 \> \< \v A + \v b_1 |,
\label{29}
\eea
%


while the kinetic operators $\Pi, \Pi^{\dagger}$ are defined as

\bea
\Pi = \sum_{\v A, i} \Delta \left( |\v A \> \< \v A + \v b_i - \v b_1 | + |\v A + \v b_1 \> \< \v A + \v b_i | \right).
\label{29.1}
\eea
The spectrum and eigenfunctions are obtained again by squaring $H$. With $M$ and $E_0$ given as before, we obtain

\bea
H= E_0 + M\sigma_3 + \sigma_+ \Pi + \sigma_- \Pi^{\dagger}
\label{33}
\eea
and

\bea
(H-E_0)^2 = M^2 + \Pi \Pi^{\dagger}
\label{33.1}
\eea

\begin{figure}[h]
\begin{center}
\includegraphics[scale=0.6]{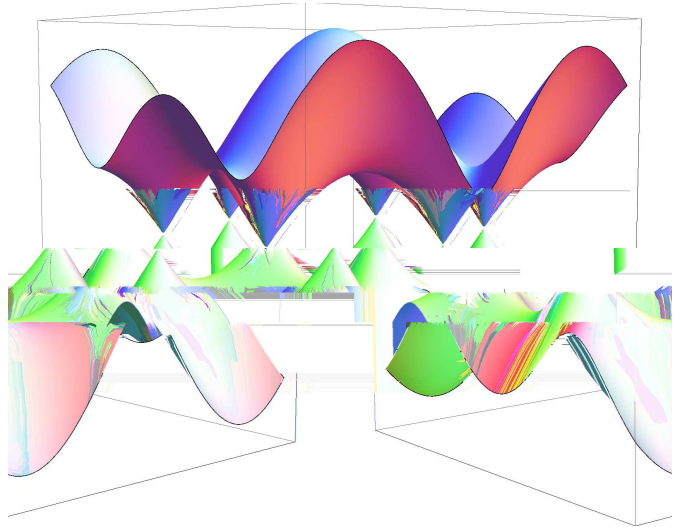}
\end{center}
\caption{Energy surfaces computed form our dispersion formula in a periodic lattice. The conical points are visible at the six corners of the Brillouin zone.}
\label{cones}
\end{figure}

The spectrum and eigenfunctions are then

\bea
E(\v k)= E_0 \pm \sqrt{\Delta^2 |\sum_i e^{i2 \pi \lambda \v b_i \cdot \v k}|^2 + M^2 } 
\label{35}
\eea
\bea
\psi^{\pm} = C^{\pm} \phi^{1}_k + D^{\pm} \phi^{2}_k, \qquad C^{\pm} = \frac{\pm(E(\v k)-E_0)-M}{\Delta (\sum_i e^{i2 \pi \lambda \v b_i \cdot \v k})} D^{\pm}
\label{36} 
\eea
It is well known that the degeneracy points of the spectrum for the massless case are $\v k_0 = \pm \frac{1}{2\lambda} (1,-\sqrt{3})$. Around such points one finds

\bea
E(\v k - \v k_0) - E_0 = \pm \sqrt{\Delta^2 k^2 + M^2}
\label{37}
\eea

\subsubsection{Tight binding and approximate isotropy}


We claim that rotational symmetry around conical points is a direct consequence
of the tight binding approximation, as we shall see. It is well known that
rotational symmetry in the Dirac equation demands a transformation of both orbital
and spinorial degrees of freedom. It is in the orbital part that we shall concentrate
by studying the energy surfaces around degeneracy points beyond the tight binding model.
In our study, it will suffice to look inside the first Brillouin zone since the rest
of the reciprocal lattice can be obtained by periodicity. Small deviations from
degeneracy points (denoted by $\v k_0$) in the form $\v k =\v k_0 + \bfkappa$ give the energy

\bea
E= \Delta | \sum_{i} \exp{(i  \lambda (\v k_0 + \bfkappa) \cdot \v b_i)} | \simeq \Delta \lambda|\bfkappa |,
\label{46}
\eea
which is rotationally invariant in $\bfkappa$.


A second-neighbor interaction of strength $\Delta'$ modifies the kinetic operator $\Pi$ as

\bea
\Pi= \Delta \sum_{i=1,2,3} T_{\v b_i} + \Delta' \sum_{i=1,2,3} T_{\v a_i} + T_{-\v a_i},
\label{47}
\eea
where the vectors $\v a_i$ have now appeared, connecting a point with its six second neighbors. The energy equation becomes

\bea
E= | \Delta \sum_{i} \exp{(i \lambda \v k \cdot \v b_i)} + \Delta' \sum_{i} 2\cos{( \lambda \v k \cdot \v a_i)} |.
\label{48}
\eea

We expect a deviation of degeneracy points $\v k'_0$, for which $\v k = \v k'_0 + \bfkappa$. Upon linearization of the exponentials in $\bfkappa$ we find the energy

\bea
E \simeq \sqrt{ (\bfkappa \cdot \v u)^2 + (\bfkappa \cdot \v v)^2}
\label{49}
\eea
where the vectors are given by

\bea
\v u = \lambda \Delta \sum_i \cos( \lambda \v k'_0 \cdot \v b_i) \v b_i  \\
\v v = \lambda \Delta \sum_i \sin( \lambda \v k'_0 \cdot \v b_i) \v b_i + 2 \lambda \Delta' \sum_i \sin( \lambda \v k'_0 \cdot \v a_i) \v a_i
\label{50}
\eea

\begin{figure}
\begin{center}
\includegraphics[scale=0.6]{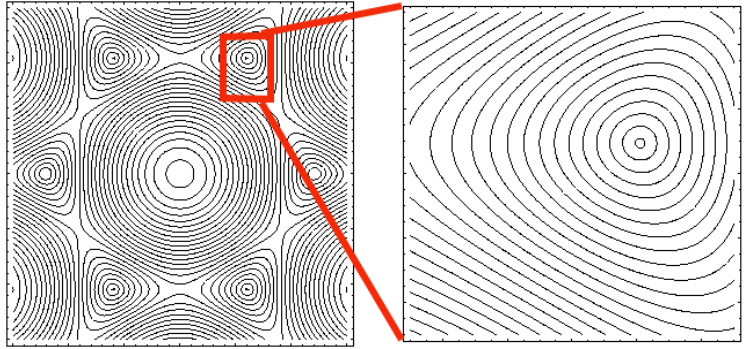}
\end{center}
\caption{First neighbor interaction, circular contours near the corners of the first Brillouin zone}
\label{contours1}
\end{figure}

\begin{figure}
\begin{center}
\includegraphics[scale=0.6]{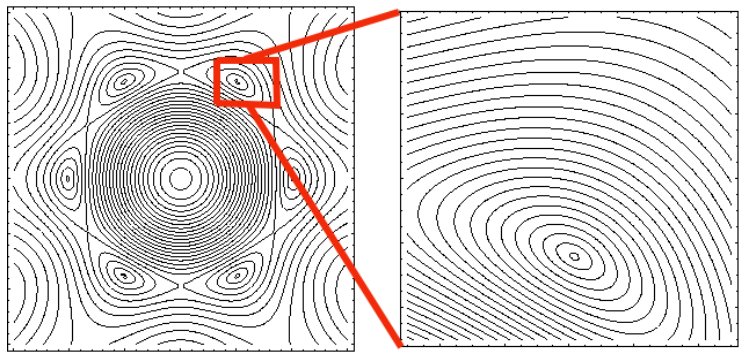}
\end{center}
\caption{Second neighbor interaction, elliptic contours}
\label{contours2}
\end{figure}

The presence of $\Delta'$ gives the energy surfaces (\ref{49}) as cones
with elliptic sections whenever $\bfkappa$ is inside the first Brillouin zone.
Regardless of how we complete the energy contours to recover periodicity,
it is evident that the resulting surfaces are not invariant under rotations 
around degeneracy points. The circular case is recovered only when $\Delta'=0$,
leading to $\v k'_0 = \v k_0$. In this case, the vectors reduce to
$\v v = (1,0),\v u = (0,1)$ when $\v k_0$ is the degeneracy point at $(1/2\lambda,0)$.

In summary, extending the interactions to second neighbors has the effect of
breaking the isotropy of space AROUND CONICAL POINTS, which is an
essential property of the free Dirac theory.

 \subsection{The Dirac oscillator in 2D}


We deform the lattice through an extension of the kinetic operators,
just as in the one dimensional case. 
Let us consider site dependent transition amplitudes $\Delta(\v A, \v A + \v b_1)$
connecting the sites labeled by $\v A, \v A + \v b_1$. Again, 
these are related to distances $d(\v A, \v A + \v b_1)$ between resonators as 
$\Delta(\v A, \v A + \v b_1)=\Delta \exp(-d(\v A, \v A + \v b_1)/\Lambda)$. 
Now we define the ladder operator

\bea
a_r= \sum_{\v A, i} \Delta(\v A, \v A + \v b_i) \left( |\v A \> \< \v A + \v b_i - \v b_1 | + |\v A + \v b_1 \> \< \v A + \v b_i | \right)
\label{38}
\eea
and impose $[a_r,a_r^{\dagger}]=\omega$. After some algebra, one can prove that this leads to three recurrence relations. The first relation is

\bea
 \Delta(\v A, \v A + \v b_1) = \Delta,
\label{39.1}
\eea
meaning that the vertical distances are fixed as a constant (the coupling is a constant $\Delta$). The second and third relations give

\bea
 \Delta^2(\v A, \v A + \v b_2) + \Delta^2(\v A + \v b_2, \v A + \v b_2 - \v b_3) = \\ \nonumber
\Delta^2(\v A + \v b_1, \v A + \v b_1 + \v b_2) + \Delta^2(\v A + \v b_2 + \v b_1, \v A + \v b_1 + \v b_2 - \v b_3),
\label{39.2}
\eea

\bea
 \Delta^2(\v A, \v A + \v b_2) + \Delta^2(\v A, \v A + \v b_3) = \\ \nonumber
\Delta^2(\v A + \v b_1, \v A + \v b_1 - \v b_3) + \Delta^2(\v A + \v b_1, \v A + \v b_1 - \v b_2) + \omega. 
\label{39.3}
\eea
It is the third relation what gives the scaling of distances in terms of our frequency $\omega$: distances should increase in order to satisfy (\ref{39.3}). The second relation simply establishes the equality of the lengths of opposite sides for each hexagonal cell.
These relations seem to be complicated, but one can use a program to generate all lattice points consistently. We do so by starting with a regular hexagonal cell. The analogy between our model and the 2 dimensional Dirac oscillator becomes exact when the number of resonators is large.

\begin{figure}[h]
\begin{center}
\includegraphics[scale=0.6]{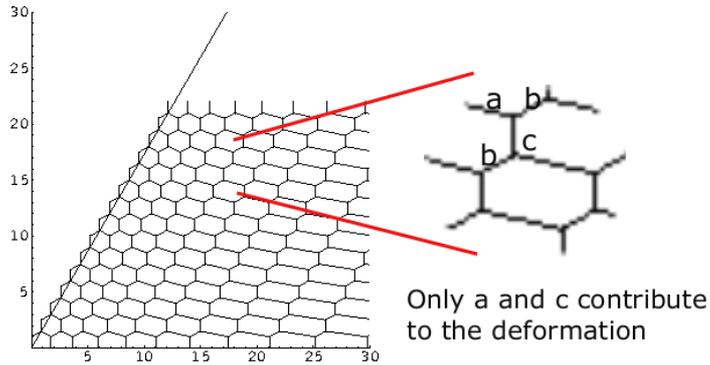}
\end{center}
\caption{A lattice produced with our recurrence relation. A regular hexagonal cell is used as
a seed. A resonator is placed on each vertex of the lattice. A choice of deformation angle may produce periodicity in one direction (trivial case)}
\label{case1}
\end{figure}



\begin{figure}[h]
\begin{center}
\includegraphics[scale=0.6]{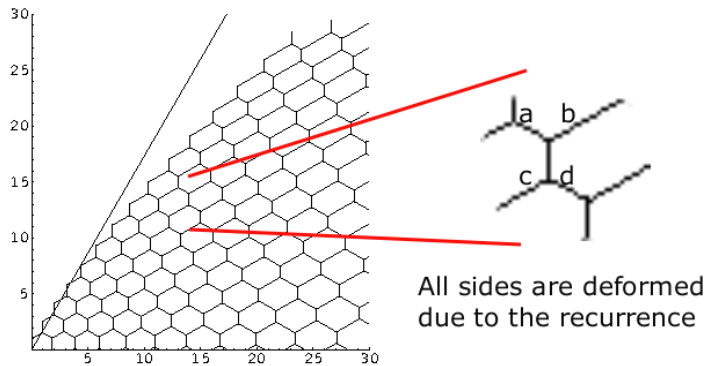}
\end{center}
\caption{A lattice produced with our recurrence relation. Resonators are placed at the vertices of the array. A regular hexagonal cell is used as
a seed (at the origin). No periodicity.}
\label{case4}
\end{figure}

The resulting hamiltonian of this problem is

\bea
H= E_0 + \sigma_3 M + \sigma_+ a_r + \sigma_- a_r^{\dagger}
\label{42}
\eea
with eigenvalues

\bea
E(N_R)= E_0 \pm \sqrt{ \omega (N_r+1) + M^2 }, \qquad 0 < N_r < \Delta^2/\omega
\label{43}
\eea
The results obtained so far confirm our suspicion that the spectrum around conical points becomes more spaced with a square root law. See the figures for reflection and transmission measurements between antennas in the array. The peaks are localized around the blue cone located at the resonance, where the Dirac point should lie.

\begin{figure}
\begin{center}
\includegraphics[scale=0.5]{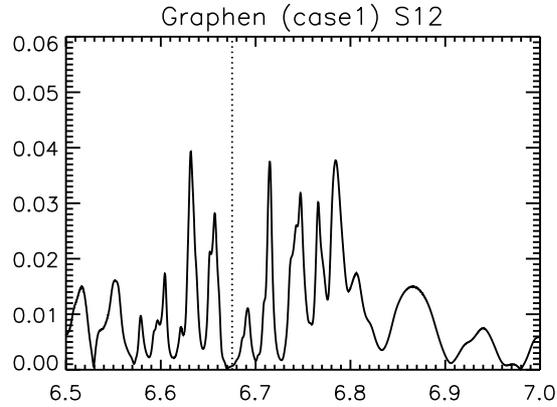}
\end{center}
\caption{Preliminary experimental results for the transmission between antennas in the array of figure \ref{case1}, as a function of the frequency (GHz). The blue line indicates the Dirac point. The
equally spaced spectrum appears due to the deformation. The gap indicates the
zero point energy of the oscillator. Courtesy of F. Mortessagne.}
\label{spectrumcase1}
\end{figure}

\begin{figure}
\begin{center}
\includegraphics[scale=0.5]{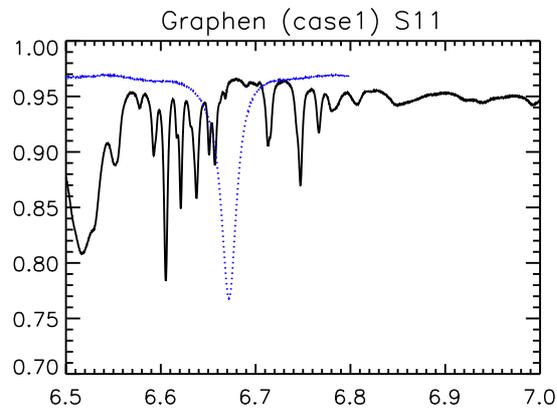}
\end{center}
\caption{Preliminary experimental results for the reflection in the array of figure \ref{case1}, as a function of the frequency (GHz). The spaced spectrum appears due to the deformation (similar to a square root law, but with asymmetries). A blue curve indicates the location of the Dirac point. The gap indicates the
zero point energy of the oscillator. Courtesy of F. Mortessagne}
\label{reflexcase1}
\end{figure}

Summarizing the results of this section, we have formulated a Dirac equation in hexagonal lattices and justified the use of tight binding arrays, together with an experimental evidence of nearest neighbor coupling and its exponential decay as a function of separation distance.
We provided a useful description for a problem motivated by graphene
and the emulation of Dirac-Moshinsky oscillators in electromagnetic billiards. Moreover, we have developed a method to analyze deformations through the algebraic properties of the system, an idea that opens a window for the realization of other integrable systems. The experimental realization of the DMO depends crucially on the measured reflection peaks, as shown in the preliminary experimental results in the figure. So far, the location of Dirac points has been successful \cite{kuhl} and the distortion of the spectrum upon deformations is also visible. It is left to run more experiments
in order to have a clear indication of a square root law for the spectrum and a localization of wavefunctions provided by the constantly increasing distance between resonators. It must be mentioned that a large number of such resonators in our setup is mandatory, allowing the possibility of neglecting finite-size and boundary effects.

\section{Conclusion}

We have reviewed the subject starting from its simple formulation as a potential problem, then revisited some of its achievements in hadron spectroscopy and finally gave the reasons why this system can be realized in nature by careful constructions. The original system proposed by Moshinsky is simple enough to be considered as a paradigmatic model. Yet it possesses a richness of interpretations provided by its many formulations (in many dimensions and for many particles), by its original applicability to bound and composite systems and, in recent times, by the analogies that can be established in connection with two areas that are active and prolific: two dimensional materials and quantum-optical traps. 

\section*{Acknowledgments}

The author is grateful to the organizers of the ELAF 2010 for their kind hospitality.

\end{document}